\RequirePackage{fix-cm}
\documentclass[smallextended]{svjour3}       
\smartqed  
\usepackage{graphicx}
\usepackage{rotating} 

\usepackage{color,soul}
\usepackage{adjustbox}
\begin{document}

\title{Chemical and Isotopic Composition Measurements on Atmospheric Probes Exploring Uranus and Neptune}



\author{Audrey Vorburger \and
        Peter Wurz \and
        Hunter Waite
}


\institute{A. Vorburger \at
              Physics Institute, University of Bern, Bern, Switzerland. \\
              Tel.: +41 31 631 44 16\\
              \email{vorburger@space.unibe.ch}
           \and
           P. Wurz \at
              Physics Institute, University of Bern, Bern, Switzerland.
           \and
           H. Waite \at
           	  Southwest Research Institute, Space Science and Engineering Division, San Antonio, Texas, United States of America.
}

\date{Received: date / Accepted: date}

\maketitle

\begin{abstract}

So far no designated mission to either of the two ice giants, Uranus and Neptune, exists. Almost all of our gathered information on these planets comes from remote sensing. In recent years, NASA and ESA have started planning for future mission to Uranus and Neptune, with both agencies focusing their attention on orbiters and atmospheric probes. Whereas information provided by remote sensing is undoubtedly highly valuable, remote sensing of planetary atmospheres also presents some shortcomings, most of which can be overcome by mass spectrometers. In most studies presented to date a mass spectrometer experiment is thus a favored science instrument for {\it in situ} composition measurements on an atmospheric probe. Mass spectrometric measurements can provide unique scientific data, i.e., sensitive and quantitative measurements of the chemical composition of the atmosphere, including isotopic, elemental, and molecular abundances. In this review paper we present the technical aspects of mass spectrometry relevant to atmospheric probes. This includes the individual components that make up mass spectrometers and possible implementation choices for each of these components. We then give an overview of mass spectrometers that were sent to space with the intent of probing planetary atmospheres, and discuss three instruments, the heritage of which is especially relevant to Uranus and Neptune probes, in detail. The main part of this paper presents the current state-of-art in mass spectrometry intended for atmospheric probe. Finally, we present a possible descent probe implementation in detail, including measurement phases and associated expected accuracies for selected species.


\end{abstract}

\section{Introduction}
\label{sec:Introduction}
The outer planets Jupiter, Saturn, Uranus, and Neptune were key players during the formation and evolution of our Solar System. They contain over 99.5\% of the planetary mass and most of the angular momentum of our Solar System. Having been formed from the same material that constitutes the proto-Sun and the surrounding proto-planetary disk, their chemical composition gives valuable insight into the formation and evolution history of our Solar System (e.g., \cite{Mousis2014} and references therein). 

Direct access to the bulk composition of the giant planets is impossible, though. Instead, their bulk compositions have to be deduced from the planets' mean densities and their gravity fields, from the abundances of the atmospheric constituents, assisted by modeling efforts. Species that hold key information on the formation and evolution process of our Solar System include the major volatiles CH$_4$, CO, NH$_3$, N$_2$; the noble gases He, Ne, Ar, Kr, Xe; and the isotopic ratios D/H, $^{13}$C/$^{12}$C, $^{15}$N/$^{14}$N, $^{3}$He/$^{4}$He, $^{20}$Ne/$^{22}$Ne, $^{38}$Ar/$^{36}$Ar, $^{36}$Ar/$^{40}$Ar, as well as those of Kr and Xe, e.g., \cite{Mousis2014,Atreya2018,Atreya2020,Cavalie2020,Simon2020}. Some of these species, e.g. the noble gases, are only present in extremely small quantities in the atmosphere, though. To make matters worse, their abundance has to be determined with extremely high accuracy (on the 1\% level and even better for the noble gases). Aggravating this situation is the fact many species are locked away in a condensed phase beneath clouds and haze layers, necessitating the capability to measure them locally. Moreover, noble gases can only be measured {\it in situ}. 

Instruments designed to probe atmospheric samples accordingly have to fulfill a wide variety of requirements. From an analytical performance point of view they have to maximize sensitivity, selectivity, throughput, precision, and accuracy, and minimize cycle times. Due to logistical constraints the instruments also have to be as small, light, and energy-efficient as possible. In addition, such instruments have to withstand harsh environments (including damaging radiation and corrosive effects), large variations in temperature and pressure, and high G-forces, micro-gravity conditions, as well as high shocks and vibrations (during take-off and atmosphere entry). Finally, their inaccessibility once launched necessitates highly reliable operation, advanced automation, and redundancy.

Atmospheric probing with mass spectrometers offers a solution to many of the challenges mentioned above. Mass spectrometers sample atmospheric species {\it in situ} with high sensitivity, dynamic range, resolution, and accuracy, providing the capability of identifying atmospheric constituents across a wide range of masses and concentrations. Today's state-of-the-art mass spectrometers often fall into one of two categories: stand-alone high-resolution mass spectrometers and low-resolution mass spectrometers coupled with additional subsystems. Novel high-resolution mass spectrometers actually used in space reach mass resolutions, $m/\Delta m$, of up to 9,000 \cite{Balsiger2007,Scherer2006}. Laboratory prototype mass spectrometers developed for space research reach mass resolutions up to several 100,000, e.g., \cite{Avice2019,Briois2016,Brockwell2016,Darrach2015,Horst2012,Hu2005,Madzunkov2014,Makarov2000,Nikolic2019,Okumura2004,Toyoda2003}. Such instruments are capable of determining isotopic ratios at the 1\% accuracy level, though at the expense of mass range. These instruments are quite complex in operation, though, and they are large, heavy and power-consuming. Low resolution mass spectrometers are simpler, smaller, lighter, and consume less power than their high-resolution counterparts, but have to be coupled with additional subsystems (e.g., for chemical pre-separation to be discussed below) to avoid isobaric interferences in the mass spectra, e.g., \cite{Mahaffy2012,Niemann2002,Niemann2005,Niemann1996,Niemann1998,Niemann2010,Niemann1992}. Isobaric interferences are caused by species having the same nominal mass, but differ slightly in their actual mass, thus they can only be separated if the mass resolution is sufficiently high. For example separating CO from N$_2$, both having nominal mass of 28, needs a mass resolution of about 3000 to actually separate them in a mass spectrum. 

In this paper we give an overview over mass spectrometry dealing with gaseous samples (Section~\ref{sec:Mass_Spectrometry}), give a short introduction into the history of atmospheric mass spectrometers flown during past space missions (Section~\ref{sec:History}), present three mass spectrometers with particularly impactful heritage for possible future descent probe missions (Section~\ref{sec:Heritage}), talk about the current state-of-art of atmospheric mass spectrometers suitable for space missions (Section~\ref{sec:State_of_Art}), and finally present a possible descent probe mass spectrometer implementation (Section~\ref{sec:Recommendation}).

\section{Mass Spectrometry}
\label{sec:Mass_Spectrometry}
Mass spectrometry is an analytical technique that identifies and quantifies the atoms and molecules present in a given sample. The mass spectrometer itself is part of a larger system that also contains a sample introduction system, a sample preparation system, an elaborate vacuum system, and associated electronics and computer. In the following we will limit the discussion to mass spectrometer systems that deal with gaseous samples, i.e.,  systems that will have relevance for an atmospheric probe. 

The sample introduction system ensures that the mass spectrometer's high vacuum requirements are met by only admitting gas at a suitable pressure. For atmospheric probes this is often accomplished through either direct inlets (for rarefied atmospheres) or through glass capillaries (for denser atmospheres). The sample preparation system's task is to remove unwanted elements and to enrich trace elements, to meet the scientific requirements as best as possible. The vacuum system contains valves for controlling the gas flow, pumps for maintaining vacuum conditions, chemical getters for removing selected species, temperature and pressure sensors for monitoring environmental conditions, and tubing for gas flow passage. Finally, associated electronics and computer are required to operate and control the instrument, for data acquisition and for data processing.

\begin{figure}
\includegraphics[width=1.00\textwidth]{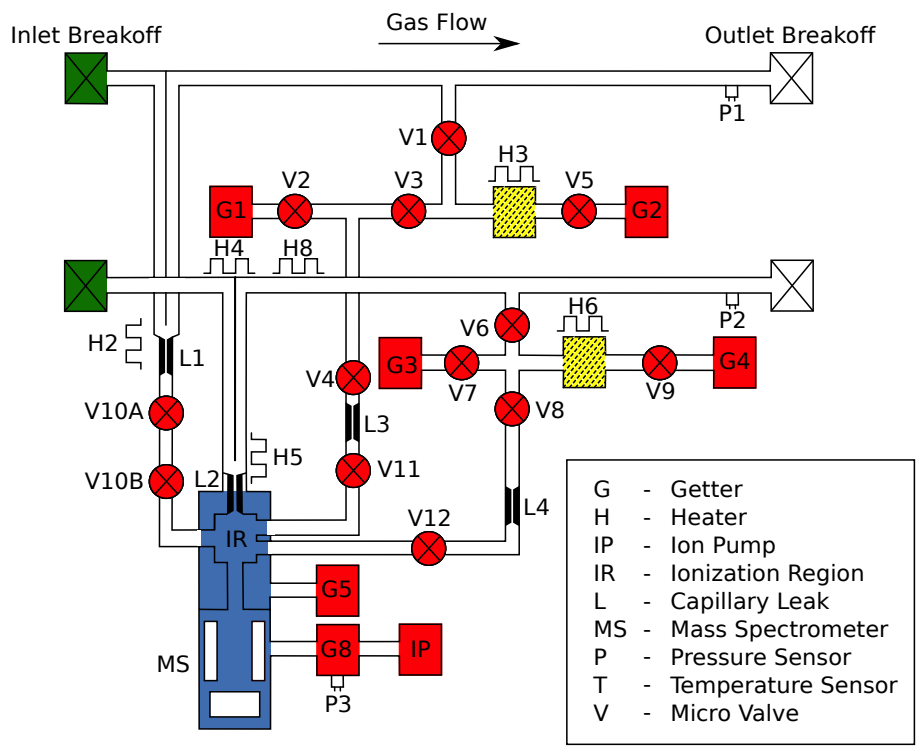}
\caption{Schematic diagram of the Galileo probe mass spectrometer system. The sample introduction system is highlighted in green, the sample preparation system is highlighted in yellow, the vacuum system is highlighted in red, and the mass spectrometer is highlighted in blue. After \cite{Niemann1996}.}
\label{fig:GPMS_Schematic}
\end{figure}

Figure~\ref{fig:GPMS_Schematic} shows a highlighted schematic of the Galileo probe mass spectrometer as an example of a mass spectrometer system. In this figure, the introduction system is highlighted in green and consists of two fully self-contained inlets all the way to their exhausts. Each of the inlets was designed for a specific pressure range, and were accordingly operated in sequence during the descent. The sample preparation system, consisting of two enrichment cells, one designed to enrich heavy noble gases and one designed to enrich complex hydrocarbons, is highlighted in yellow. The vacuum system with its valves, pumps, getters and sensors is highlighted in red. Finally, the actual mass spectrometer is highlighted in blue.

The mass spectrometer itself consists of four main components that are responsible for (i) converting the sample into gaseous ions, for (ii) separating the ions according to their mass-to-charge ratio, $m$/$z$,  for (iii) recording the separated ions, and (iv) a suitable  vacuum environment for the mass spectrometer. Moreover, often some additional subsystems are added to increase the performance of the mass spectrometer system. For example, getters, chemical and cryogenic enrichment cells,  or gas-chromatography are useful in this respect. Getters can be used to filter out certain reactive species, and are described in Section~\ref{sec:Getters}. To separate or eliminate isobaric interferences, mainly two techniques have asserted themselves in the past couple of years. These two techniques, chemical and cryogenic enrichment, are described in Subsections~\ref{sec:Chemical_Enrichment} and \ref{sec:Cryogenic_Enrichment}. Finally, low-resolution mass spectrometers can be coupled with other instrumental analysis techniques, e.g., gas chromatography (see Subsection~\ref{sec:Gas_Chromatography}), sample cleaning (see Subsection~\ref{sec:Getters}), and enrichment cells (see Subsection~\ref{sec:Chemical_Enrichment}).

\subsection{Ion source}
\label{Ion_Source}
The ion source is responsible for ionizing the neutral atoms and molecules that comprise the sample to be analyzed since mass spectrometric separation is always for ionized species. The efficiency and quality of this step is crucial for the performance of the mass spectrometer. Only if a sufficiently large fraction of the particles are ionized can atoms and molecules that are very rare in the given sample be detected. In addition, only if the degree of ionization is known from calibration, and the ionization is stable over the operation time of the instrument,  it is possible to quantify the relative abundances as well as the absolute densities of the different species. Most mass spectrometers designed for space application use electron ionization providing positive ions for analysis, in particular for gaseous samples, due to the method's simplicity and robustness. Electron ionization is a technique that dates back to the 1940s \cite{Nier1947}. However, electron ionization is a `hard ionization' technique that introduces a large degree of fragmentation of complex molecules, which effectively limits the mass range of such instruments. 

Other ionization methods are also emerging, one example of which is laser ionization in combination with laser ablation or laser desorption used for solid samples \cite{Riedo2016,Riedo2019}. In addition, the recent desire to look for large organic compounds, especially in the context of biological matter, has shifted interest  towards `soft ionization' techniques, e.g., laser desorption \cite{Goesmann2007,Riedo2015,Moreno2016,Li2017,Ligterink2019}, laser post-ionization \cite{Getty2012}, and chemical ionization \cite{Waller2019}, which introduce  less fragmentation than electron ionization. Some of these new instrument developments use high resolution mass analyzers \cite{Briois2016,Selliez2019}. 

\subsection{Mass analyzer}
\label{Mass_Analyser}
The mass analyzer separates the gaseous ions according to their mass-to-charge ratio ($m$/$z$). Mass separation can occur in a number of different ways, with the underlying operating principle giving the mass spectrometer class its name. In the past 50 years, four classes of mass spectrometers have been used for the measurement of gaseous atmospheres in space: magnetic sector mass spectrometers, e.g., \cite{Balsiger1971,Rushneck1978,Moor1989}, quadrupole mass spectrometers \cite[e.g.]{Dawson1976,March1989}, ion trap mass spectrometers \cite{Wright2007,Goesmann2007}, and time-of-flight mass spectrometers \cite{Johnson1955,Balsiger2007,Scherer2006}. Recently, also orbitrap mass spectrometers have been proposed for space application \cite{Briois2016,Selliez2019}.

\textbf{Magnetic sector} instruments introduce ions into a flight tube with a fixed bending radius and a variable magnetic field oriented perpendicular to the radial path. According to Newton's second law and the Lorentz force law, for a given field strength and a given radius, only ions with a certain $m$/$z$ value can pass through the analyzer section and reach the detector system \cite{Benninghoven1987}. All other ions travel on trajectories intersecting with the tube walls, prohibiting them from reaching the detector system. Varying the applied magnetic field strength, or the ion energy, allows successive registration of ions with different $m$/$z$ values. To collect an entire mass spectrum a scan has to be performed, which makes magnetic sector instrument slow in collecting mass spectra. Moreover, the magnet limits the largest mass that can be measured by such an instrument. Simple magnetic sector instruments have low mass resolution. Only the combination of the magnetic sector with an electric sector, the so called double focusing instruments, allows for high mass resolution.

\textbf{Quadrupole} mass analyzers separate ions by guiding them through an electric quadrupole field, created by applying a radio frequency and a direct current voltage signal to four cylindrical or hyperbolic rod electrodes, positioned parallel and equidistantly from a central axis extending in the $z$-direction \cite{Dawson1976}. Similar to magnetic mass spectrometers, for a certain voltage setting only ions with a certain $m$/$z$ value can pass through the quadrupole field in a stable manner to reach the detector. Other ions have unstable trajectories and will collide with the electrodes, never reaching the detector. Typically, quadrupole mass spectrometers have unit mass resolution, i.e., they can separate each integer mass from the next one. Moreover, since the needed high voltages for the rod system scale with mass number, there is a limit in maximum $m$/$z$ of such an instrument. Again, a full mass spectrum is obtained by successively scanning across the desired range of $m$/$z$ values, and the resulting duty cycle has a direct effect on the sensitivity of the instrument. 

\textbf{Ion trap} mass spectrometers encompass both two-dimensional (linear) and three-dimensional (quadrupole and cylindrical) ion trap mass spectrometers. They all operate on the basis of ions being confined by direct current and radio frequency electric fields in so-called `ion traps' \cite{March1989}. Trapped ions can be sequentially ejected from the trap as a function of their $m$/$z$ values to record a mass spectrum. This function can also be used clean the ion sample in the trap or enrich the sample in certain ions of choice from a gas sample collected in the ion trap. This capability allows for a certain level of pre-selection and preparation of a sample in the ion trap, which is also known as tandem mass spectrometry.  Since a typical ion trap has a maximum capacity of about 10$^5$ to 10$^6$ ions it can store \cite{March1989}, there are limits to this pre-selection and preparation of a sample when it comes to trace species at the ppm-level, and associated isotopes. 

\textbf{Time-of-flight} mass spectrometers operate on the principle that ions with the same kinetic energy are dispersed in time during their flight in a long field-free drift tube due to their different $m$/$z$ values (i.e., ions with smaller $m$/$z$ values will arrive earlier at the detector than ions with higher $m$/$z$ values) \cite{Cotter1992}. The different arrival times of different ions thus give their $m$/$z$ values. Unlike other mass spectrometer classes, time-of-flight mass spectrometers do not need to scan across a mass range to obtain a complete mass spectrum, but rather obtain it `in one go'. They also stand out by theoretically not being limited in mass range. In practice, the mass range is limited solely by the size of the signal accumulation memory. 

\textbf{Orbitrap}  mass spectrometers, which are a novel version of the Fourier transform mass spectrometer, were recently introduced \cite{Makarov2000,Hu2005}, and were proposed for space flight \cite{Briois2016}. The orbitrap mass spectrometer contains an inner and an outer electrode, the first of which resembles a spindle and the second of which resembles a barrel. Ions that enter the electrostatic field created between the two electrodes are trapped in an orbital motion around the inner spindle-electrode. During its orbit around the spindle the ion oscillates forth and back along the spindle's central axis, with an axial oscillation frequency solely dependent on the ion's mass-to-charge ratio $m$/$z$ and the trapping potential established by the inner and outer electrodes. The axial oscillation induces an image current in the outer electrode, which has two segments, that can be detected by the detector system. Multiple ions generate multiple signals, the frequencies and relative abundances of which can be analyzed using Fourier transformations. A typical orbitrap has a maximum capacity of about 10$^4$ to 10$^5$ ions it can store \cite{Grinfeld2019} before the mass resolution degrades and minor species are masked by major ions in the trap, limiting to the dynamic range. When it comes to trace species at the ppm-level, and associated isotopes suitable pre-selection or enrichment of the species of interest will be necessary. 

\textbf{Fourier transform} mass spectrometers  have been used in the laboratory for high to ultra-high resolution mass spectrometry investigations with mass resolutions exceeding 10,000,000 \cite{Marshall1990,Buchanan1987,Nikolaev2011}. Since these instruments need strong and very stable magnetic fields, usually provided by super-conducting magnet systems, e.g. \cite{Smith2018}, such instruments have not been considered for use in space flight so far.   

\subsection{Detector system}
\label{Detectors}
The last stage, the detector, converts ion intensities (total number of ions per mass, or fluxes) into an electrical signal that can then be handled by the data acquisition system. Detectors need to keep electronic background signals as low as possible, because these can `mask' peaks of rare species. Common detector types used include Faraday cups, electron multipliers, micro-channel plates, channeltrons, photo-multipliers, and ion-to-photon detectors.

\subsection{Additional Components}
\label{sec:Additional}

\subsubsection{Getters}
\label{sec:Getters}
While getters come as physical and chemical getters, chemical getters are preferably used in combination with mass spectrometers flown in space because of their simple use and long history in laboratory systems. Chemical getters contain a material that is able to fix gas molecules on their surface in the form of stable chemical compounds. If the getters retain the species that they have filtered out indefinitely, they effectively function as pumps.

All three mass spectrometers presented in Section~4 as heritage instruments implemented a getter in some form. The Galileo Probe Mass Spectrometer (GPMS)  used chemical getters to remove active gases from the sample, such as the major atmospheric species H$_2$ \cite{Mahaffy2000}, the Gas Chromatograph Mass Spectrometer (GCMS) on Huygens used getters to pump reactive gases with the exception of methane \cite{Niemann2010}, and \cite{Balsiger2007} mentions that whereas no actual chemical getters were implemented for ROSINA, the large amounts of titanium elements used for the instrument enclosure acted as an active getter surface themselves.

\subsubsection{Chemical Enrichment}
\label{sec:Chemical_Enrichment}
Chemical enrichment cells contain chemical getters that adsorb reactive gases with the goal of either improving the measurement of trace gases by removing the major components, or with the goal of removing interferences of overlapping ions. Subsequent heating of the getter desorbs the gases trapped on the cell material, allowing it to flow towards the ion source. Reference \cite{Niemann1992} shows in Figure~2 an illustrative comparison between two mass spectra, one recorded before and one recorded after enrichment. The enrichment factor depends on the chemical and physical properties of the species to be investigated, and generally increases with the molecular weight of the species. Both GPMS on Galileo and GCMS on Huygens contained enrichment cells allowing for enrichment of hydrocarbons by a factor of 100 (ethane and propane) to 500 (for higher order hydrocarbons), and for enrichment of heavy noble gases (krypton and xenon) by a factor of 10--100 \cite{Niemann1992,Niemann2002}.

\subsubsection{Cryogenic Enrichment}
\label{sec:Cryogenic_Enrichment}

Cryogenic enrichment using cryogenic coolers (cryocoolers) have been used for enrichment in laboratory-based sampling systems for many years. This technique has been applied in prototype instruments for space applications to increase the sensitivity of a mass spectrometer by a factor of up to 10,000, which was demonstrated for Ar by collecting for 500~s, see Figure 3 in \cite{Brockwell2016}. A Ricor 508 cryocooler has been foreseen for the MASPEX instrument to provide a ten-thousandfold enhancement in the gases analyzed as the instrument passes through the atmosphere of Europa during the flybys. The absorber  uses a 316L stainless steel absorber for the trapping of gases. Similar methodology could be used in entry probes to increase the signal to noise ratio of Ar, Kr, and Xe. Combining this technique with a non-evaporable getter as a primary stage to remove active volatiles would be a powerful means of making accurate measurements of Ar, Kr, and Xe and their isotopes on an atmospheric probe mass spectrometer system.

\subsubsection{Noble Gas Enrichment}
\label{sec:Noble_Gas_Enrichment}
The noble gas enhancement can be achieved by using a combination of a cryotrap, an ion pump, and a non-evaporable getter (NEG: SAES 172). The NEG removes all constituents except methane and the noble gases. The cryotrap traps the products of the NEG process, except for helium and some neon. The ion pump then operates to pump away the helium, which is the second highest source of gas in the atmosphere of a giant planet, thus enhancing the remaining noble gases about 200 times. Helium and neon are measured using a separate mode. 

\subsubsection{Gas Chromatography}
\label{sec:Gas_Chromatography}
Gas Chromatography is an established laboratory technique in analytical chemistry to separate a complex gas mixture that can be vaporized without decomposition \cite{Poole2012}. The process traditionally involves a single 10--30 meter long capillary column that is  coated on its walls with a selected stationary phase specific to the compounds of interest. An inert carrier gas (most often H$_2$ or He), the mobile phase, pushes the sample gas mixture through the column whereby allowing the species to interact with the stationary phase coating on the wall of the column. The passage time of a species through the column  depends on the chemical and physical properties and its interaction with a specific column coating, leading to the separation at the exit of the column, and allows for both qualitative and quantitative studies to be performed. Typically, the column is heated to perform a boiling point separation of the mixture, improve peak shape and reduce the overall analysis time. 

\subsubsection{Reference Gas}
\label{sec:Reference_Gas}
A gas reference system is needed to allow determination of the detector gain, especially after an 8--10 years of cruise to a giant planet, to provide a known reference for isotope measurements, and to provide an accurate mass scale for high resolution mass spectrometers. Reference gas systems were used by ROSINA for detector gain, isotope reference, and mass scale determination in DFMS and RTOF, using a gas mixture of (Ne, CO$_2$, Xe) and (He, CO$_2$, Kr), respectively \cite{Balsiger2007}. Gain determination must use a Faraday cup as well as the calibration system. The known quantity of gas from the calibrated leaks compared with the ion current provides the source sensitivity and thereby accuracy in determining the gas density. MASPEX will use a FC43 (C$_{10}$F$_{18}$) calibration gas to allow mass scale determination over the range of $m/z=$40 to 500 from the various daughter peaks produced from electron impact (EI) dissociative ionization of the calibrant.  Odd-mass EI of fragments of compounds containing F, Br, and Cl prevents isobaric interferences within the science sample in a planetary environment, and thus allows calibration and science determination to proceed in parallel to minimize calibration switching during the short probe descent time.

\section{History of Atmospheric Mass Spectrometry in Space}
\label{sec:History}
Mass spectrometers have been sent to space for over half a century (see also \cite{Palmer2001}). In the late 1950's sounding rockets carried first ion mass spectrometers into space to study the composition and number density of the lower ionosphere \cite{Bennett1950}. Whereas the mass spectrometers' sensitivities were very high, mass resolution was just high enough to allow separation of the ionospheric ions. These first ion mass spectrometers were followed by neutral mass spectrometers, the first two of which were launched into Earth's atmosphere in 1963 \cite{Nier1964}. They were both magnetic mass spectrometers: one, a double-focusing instrument, the other a single-focusing instrument. Measured atmospheric neutral species included N$_2$, O, Ar, and Ne.

The first mass spectrometers to sample the atmosphere of a Solar System body other than Earth were carried on the last three Apollo missions, Apollo 15--17. The Apollo 15 and 16 mass spectrometers were deployed from the Scientific Instrument Module on the Service Module on a 7.3-meter long boom \cite{Hoffman1972}. Most gases detected were associated with the spacecraft itself with the exception of $^{20}$Ne, which was attributed to external sources, i.e., to be of lunar origin. The mass spectrometer on Apollo 17 was deployed on the lunar surface, where it measured gas concentrations down to 10$^{-15}$~mbar, positively identifying the exospheric Ne, He, H, CH$_4$, CO$_2$, NH$_3$, H$_2$O, and $^{40}$Ar \cite{Hoffman1973}. 

In 1967, Venera 4 became the first space probe to measure the atmosphere of another planet, namely Venus \cite{Vinogradov1971}. Two gas analyzers (chemical cells) on board Venera 4 were designed to measure CO$_2$, N$_2$, O$_2$, and H$_2$O during the probe's descent by a pressure difference (manometric) method. The Venera 4 mission was followed by several successor missions (Venera 5 through 14 [1969--1982]), which sampled Venus' atmosphere from in orbit, during descent, and while stationed on the ground. The later Venera missions and the Pioneer Venus missions (1978--1992) carried more sophisticated gas analyzers (mass spectrometers and gas chromatographs) quantifying several additional elements and molecules (including the scientifically important $^3$He and HD) present in Venus' atmosphere \cite{Colin1979,Istomin1980}.

In the past 40 years, the atmospheres of many more Solar System objects have been sampled and analyzed. Mars, in particular, has seen a wide range of atmospheric mass spectrometers: From the first double-focusing magnetic mass spectrometers on Viking 1 and 2 \cite{Nier1977,Rushneck1978}, which quantified the most common species and their isotopes in the martian atmosphere in 1976, to the quadrupole mass spectrometer on MAVEN (2014--today), which measured a wide array of species and isotopic ratios \cite{Mahaffy2015}. BepiColombo, currently en route to Mercury, will complete the list of terrestrial planets the atmospheres of which have been investigated through mass spectrometry \cite{Schulz2006}. The STart from a ROtating FIeld mass spectrOmeter (STROFIO) is designed to cover a mass range of $m/z=$1--64, with a mass resolution of $m/\Delta m>$60 and a sensitivity of 0.14~(counts/sec)/(part/cm$^3$) \cite{Orsini2010}, corresponding to a signal of 10$^{-16}$~mbar measured in 10 s integration. 

So far only two missions with a mass spectrometer have been sent to analyze the atmospheres of bodies located in the outer Solar System: Galileo, the probe of which entered and analyzed Jupiter's atmosphere on 7 December 1995 \cite{Niemann1992}, and Cassini-Huygens, the probe of which landed on Titan on 14 January 2005 \cite{Niemann2002}. The mass spectrometers of both space probes are discussed in detail in the Section~\ref{sec:Heritage}. 

Finally, three comets have so far been extensively studied by atmospheric probes: Halley's comet, which was visited by the Giotto in 1986 \cite{Krankowsky1986}, Wild 2, visited by the Stardust mission in 2004 \cite{Kissel2003}, and 67P/Churyumov-Gerasi\-menko, which was accompanied by Rosetta from 2014 through 2016 \cite{Balsiger2007,Glassmeier2007}. The instrument specifications for the mass spectrometers onboard Rosetta are also presented in detail in Section~\ref{sec:Heritage}.

Planned missions for the near-future that include mass spectrometers for atmospheric investigation are, amongst others, JUICE, the JUpiter ICy moons Explorer \cite{Grasset2013}, Europa Clipper \cite{Pappalardo2013}, Comet Interceptor \cite{Snodgrass2019}, and Dragonfly. The Neutral Ion Mass spectrometer (NIM), part of the Particle Environment Package (PEP) onboard JUICE set for launch in 2022, will conduct the first-ever direct sampling of the exospheres of Europa, Ganymede, and Callisto \cite{Barabash2013}. NIM has a mass resolution $m/\Delta m$ of more than 1,100 in the mass range $m/z=$1--1,000 and a detection limit of $\sim$1~cm$^{-3}$ (corresponding to 10$^{-16}$~mbar) for a 5~s integration time. The MAss SPectrometer for Planetary EXploration (MASPEX) onboard Europa Clipper \cite{Brockwell2016} will analyze the composition of Europa's tenuous atmosphere, with a planned launch in 2025 \cite{Foust2019}. The instrument is designed to measure particles with masses up to $m/z>$1,000, exhibit an extremely high mass resolution of $m/\Delta m >$30,000, has a sensitivity of more than 1 ppt with cryotrapping and should reach a dynamic range of 10$^9$ in 1~s. Finally, the Mass Analyzer for Neutrals in a Coma (MANiaC), due to launch in 2028 onboard the Comet Interceptor spacecraft of ESA, is a mass spectrometer designed to analyze the composition of an as of yet undetermined comet's or interstellar object's gas coma. MANiaC is designed to cover a mass range of $m/z=$1--1,000, to exhibit a mass resolution of $m/\Delta m \sim$1,000, and to cover a pressure range of 10$^{-6}$--10$^{-16}$~mbar, based on the predecessor  instruments NIM for JUICE and NGMS for Luna Resurs. Dragonfly is set to launch in 2026, with the goal of reaching Saturn's largest moon Titan in 2034. Dragonfly will carry a mass spectrometer called DraMS (Dragonfly Mass Spectrometer), intended to identify chemical components, especially those relevant to biological processes, in Titan's surface and atmospheric samples. DraMS will be based on heritage from SAM (Sample Analysis at Mars) on Curiosity and MOMA (Mars Organic Molecular) on ExoMars \cite{Lorenz2018}.

Table~\ref{tab:Missions} gives an overview of missions with mass spectrometers for gas analysis that were already sent or are planned to be sent to space. The table lists the mission designations, the subsystem that carried the instrument (where available), the instrument's name, the type of mass spectrometer (magnetic, quadrupole, time-of-flight, or ion trap), if gas chromatography was applied, and the years during which atmospheric measurements were conducted.

\begin{table}[ht]
\fontsize{7pt}{7pt}\selectfont
\caption{Missions with mass spectrometers sent (planned to be sent) to space.}
\label{tab:Missions}
\begin{tabular}{lllcccc}
\hline\noalign{\smallskip}
Body of interest        & Mission                   & Subsystem                 & Instrument            & Type              & GC Addition   & Year \\
\noalign{\smallskip}\hline\noalign{\smallskip}
Mercury                 & BepiColombo               & Mercury Planet Orbiter    & STROFIO               & Time-of-flight    & No			& 2025-- \\
Venus                   & Venera 4--6               & -                         & -                     & Quadrupole        & No			&               1967--1969 \\
                        & Venera 7--14              & -                         & -                     & Quadrupole        & No			& 1969--1982 \\
                        & Pioneer Venus Orbiter     & -                         & ONMS                  & Quadrupole        & No			& 1978--1992 \\
                        & Pioneer Venus Multiprobe  & Four probes               & LNMS                  & Magnetic          & No			& 1978--1978 \\
                        & Pioneer Venus Multiprobe  & Bus                       & BNMS                  & Magnetic          & No			& 1978--1978 \\
                        & Vega 1                    & Descent craft             & ING                   & Time-of-flight    & No			& 1985--1987 \\
                        & Vega 2                    & Descent craft             & ING                   & Time-of-flight    & No			& 1985--1987 \\
Moon                    & Apollo 15                 & Orbiter                   & -                     & Magnetic          & No			& 1971--1971 \\
                        & Apollo 16                 & Orbiter                   & -                     & Magnetic          & No			& 1972--1972 \\
                        & Apollo 17                 & Lander                    & LACE                  & Magnetic          & No			& 1972--1972 \\
                        & Chandrayaan-1             & Moon Impact Probe         & CHACE                 & Quadrupole        & No			& 2008--2008 \\
                        & LADEE                     & -                         & NMS                   & Quadrupole        & No			& 2013-2014 \\
                        & Luna Resurs				& -							& GAP					& Time-of-flight	& Yes			& 2025-- \\
Mars                    & Viking 1                  & Lander                    & GCMS                  & Magnetic          & Yes			& 1976--1982 \\
                        & Viking 2                  & Lander                    & GCMS                  & Magnetic          & Yes			& 1976--1980 \\
                        & Phoenix                   & Lander                    & TEGA                  & Magnetic          & No			& 2008--2008 \\
                        & Mars Science Laboratory   & Curiosity rover           & QMS                   & Quadrupole        & Yes			& 2012-- \\
                        & MAVEN                     & -                         & NGMS                  & Quadrupole        & No			& 2014-- \\
                        & Mars Orbiter Mission      & -                         & MENCA                 & Quadrupole        & No			& 2014-- \\
                        & ExoMars                   & Rosalind Franklin rover   & MOMA                  & Ion trap          & No			& 2021-- \\
Jupiter                 & Galileo                   & Probe                     & GPMS                  & Quadrupole        & No			& 1995-1995 \\
                        & JUICE                     & -                         & NIM                   & Time-of-flight    & No			& 2030-- \\
                        & Europa Clipper            & -                         & MASPEX                & Time-of-flight    & No			& 2025-- \\
Saturn - Titan (moon)   & Cassini-Huygens           & Orbiter                   & INMS                  & Quadrupole        & No			& 2004-2017 \\
                        & Cassini-Huygens           & Probe                     & GCMS                  & Quadrupole        & Yes			& 2005-2005 \\
                        & Dragonfly                 & -                         & DraMS                 & Ion trap          & Yes			& ? \\
Comet - Halley          & Giotto                    & -                         & NMS                   & Magnetic          & No			& 1986--1986 \\
Comet - 67P             & Rosetta                   & -                         & DFMS                  & Magnetic          & No			& 2014--2016 \\
                        & Rosetta                   & -                         & RTOF                  & Time-of-flight    & No			& 2014--2016 \\
                        & Rosetta                   & -                         & COSAC                 & Time-of-flight	& Yes			& 2014--2016 \\
Comet - unknown         & Comet Interceptor         & -                         & MANIaC                & Time-of-flight    & No 			& ?\\
\noalign{\smallskip}\hline
\end{tabular}
\end{table}

\section{Heritage for Atmospheric Descent Probes}
\label{sec:Heritage}
In the following subsections we present three highly impactful and heritage-providing mass spectrometers that were successfully flown on space missions. Table~\ref{tab:GPMS_GCMS_DFMS_RTOF} lists their key properties and performances to facilitate comparison between the three instruments.

\subsection{The Galileo Probe Mass Spectrometer on the Galileo Probe}
\label{sec:GPMS}
The Galileo Probe Mass Spectrometer (GPMS) was designed to measure major and minor species present in Jupiter's atmosphere as the Galileo probe descended through Jupiter's atmosphere \cite{Niemann1992}. The probe was released into the atmosphere on 7 December 1995 and measured atmospheric species for a total of 57~minutes, starting at an ambient pressure of $\sim$500~mbar and measuring up to an ambient pressure of $\sim$20~bar \cite{Niemann1998}. The instrument, a quadrupole mass spectrometer, had a compact design, with a quadrupole field radius equal to 5.0~mm and a field length of 150~mm, weighed 13.2~kg, and required $\sim$13~W (12~W) of power for the instrument (heaters). The mass analyzer covered ions from $m/z=$2 to 150 with unit mass resolution and a maximum dynamic range of 10$^8$, allowing detection of gases with mixing ratios larger than a few tens of ppbv. The analyzer used a stepping scanning format, with step size equal to integer or 0.125~u and a measurement period of 0.5~s per mass step. A nominal scan ($m/z=$2--150 with a unit step size) thus required a scan period of 75~s. Ionization of the neutral gas was achieved through electron ionization, with a dual filament ion source (one for redundancy) with adjustable electron beam energy. The ion detector consisted of a secondary electron multiplier pulse counter.

The GPMS experiment had a quadrupole mass spectrometer with a basic gas inlet system consisting of two arrays of glass capillaries supplemented by three selective subsystems: a noble-gas purification cell and two sample enrichment cells (for more complex compounds). A schematic of the experiment is presented in Figure~\ref{fig:GPMS_Schematic}. The two basic inlet systems were fully self-contained and were operated in time sequence as the probe descended through the atmosphere. Inlet 1 was used from 520~mbar to 3.78~bar whereas inlet 2 was used from 8.21~bar to $\sim$21~bar. At certain times, while direct sampling was interrupted, the gases collected in the two enrichment cells (mainly complex hydrocarbons) were released and passed to the ionization region of the mass analyzer for analysis. The enrichment cells allowed for enrichment of hydrocarbons by a factor 100--500. The enrichment cells also each contained two getters that chemically bound reactive gases (mainly hydrogen, the major component of Jupiter's atmosphere), increasing the relative abundances of all species by approximately a factor of 10 and allowing the analysis of purified noble gases. This resulted in the measurement of isotopic ratios with better precision, including the scientifically highly important D/H ratio. Background measurements were taken when time permitted (for example in between the switching from inlet 1 to inlet 2). Finally, towards the end of the measurement sequence, high resolution scans were performed at selected masses.

During the probe's descent signals were recorded from over 6,000 individual values of mass-to-charge ratio ($m$/$z$) \cite{Niemann1996}. Species detected include H$_2$ and HD, $^{3}$He and $^{4}$He; the isotopes of the noble gases Ne, Ar, Kr, and Xe; the volatiles CH$_{4}$, NH$_{3}$, H$_{2}$O, H$_{2}$S; a chlorine compound which may have been HCl, and a large number of C$_2$ and C$_3$ non-methane hydrocarbons, but no heavier hydrocarbons \cite{Mahaffy2000,Niemann1998,Niemann1998b,Wong2004}.

The Galileo Probe Mass Spectrometer left several open questions that need to be considered in future atmospheric probe missions:
\begin{enumerate}
	\item Did the designed sampling system comprehensively sample gases, condensates, and adsorbed compounds? 
	\item Why was the highest signal to noise data collected in the enrichment cell and not in the noble gas cell?
	\item Did switching between direct sampling and sampling of the alternative collection cells lead to sampling delays that made the chronology of the probe's descent data set harder to reconstruct and understand? 
	\item How did the low mass resolution affect the interpretation of the results?
\end{enumerate}
The first point is particularly relevant given the recent Juno results on the distribution of ammonia in Jupiter's upper atmosphere inferred from the microwave opacity \cite{Bolton2017}. These measurements were interpreted as distribution of ammonia in the vapor phase as function of altitude and latitude, indicating that the distribution of volatiles and condensates is spatially dependent at Jupiter. Therefore sampling of both vapor and condensates is essential for determining the correct  mixing ratios of volatile species. Clouds are expected to be encountered during the descent of the probe in Uranus' and Neptune's atmosphere \cite{Atreya2019,Mousis2019}. 



\subsection{The Gas Chromatograph Mass Spectrometer on the Huygens Probe}
\label{sec:GCMS}
The Cassini-Huygens mission consisted of two main elements: The Cassini orbiter, which orbited Saturn for over 13 years (June 2004 -- September 2017), and the Huygens probe, which descended through to Titan's atmosphere all the way to the surface  in January 2005. The lander probe carried a Gas Chromatograph Mass Spectrometer (GCMS) experiment designed to identify and characterize the chemical composition of the atmosphere of Titan and to determine the isotope ratios of the major gaseous constituents \cite{Niemann2002}. The instrument measured atmospheric species for 148~minutes during descent (from $\sim$146~km down to the surface). Having survived the impact on the surface, GCMS measured species evaporated from the surface  for an additional 72~minutes until contact was lost with the Cassini orbiter, which relayed the communications \cite{Lebreton2005}. The GCMS was of cylindrical shape, with a diameter of 198~mm and a height of 470~mm, weighed 17.3~kg, and required on average 41~W of power. The mass analyzer was capable of measuring ions from $m/z=$2 to 141 with unit mass resolution and a mixing ratio detection limit on the order of 10$^{-8}$. Stepping was performed at unit mass steps with a single mass step taking 5~ms to acquire, resulting in a mass scan cycle of 937.5~ms. In-flight, the mass scan cycle period doubled to 1.875~s due to transmission constraints. Ionization was achieved through electron impact ionization and ions were detected by a secondary electron multiplier detection system.

The GCMS consisted of a quadrupole mass filter that could receive ions from one of five available ion sources. The first ion source was connected to a gas sampling system, where atmospheric gas was directly admitted. The second ion source was fed from an aerosol collector pyrolyzer (ACP), where aerosol samples were collected. The last three ion sources were connected to a gas chromatographic system consisting of three gas chromatographic columns (GCs). The three GC columns were designed to separate C$_3$ to C$_8$ hydrocarbons and nitriles, C$_1$ to C$_3$ hydrocarbons, and N$_2$ and CO. Unfortunately, the ion source connected to the third GC failed to operate normally at 74~km altitude, when the probe passed through a region of extensive turbulence. With the third GC column being dedicated to separate CO and N$_2$ (both with $m$/$z$ = 28), this resulted in the loss of the data from this column and the measurement of the CO concentration. The instrument also contained an enrichment cell that adsorbed trace gases (e.g., high boiling point hydrocarbons and nitriles, but no nitrogen or noble gases) for trace species enrichment, and a chemical scrubber cell for noble gas analysis. 

Similarly to the GPMS, the atmospheric inlet of GCMS fed two sets of capillary arrays with different gas conductance used in sequence to cover the wide ambient pressure range encountered during the descent. For the first $\sim$30~min (from 146~km to 65~km) atmospheric gases were induced directly into the ion source through the first leak. Simultaneously, from 77 to 75~km, gas samples for the noble gas scrubber and the sample enrichment cell were collected. From 65 to 56~km direct sampling was interrupted and the contents of first the noble gas scrubber cell and later the sample enrichment cell were analyzed. At 56~km direct sampling was resumed, using the second leak for the ion source. Concurrently, in the upper atmosphere, GC samples were collected to be analyzed later during the descent, when more time was available for analysis. In the lower atmosphere, close to the surface, GC samples were injected directly from the atmosphere, omitting storage. ACP sampling was conducted during the later stage of the descent, when the second leak was used for direct atmospheric sampling, and aerosol analysis was performed during two short periods.

The GCMS on Huygens measured the altitude profiles of N$_2$, CH$_4$, H$_2$, and $^{40}$Ar in the lower atmosphere, detected $^{22}$Ne and $^{36}$Ar in a batch sample in the noble gas enrichment cell, determined the isotope ratios for H/D, $^{14}$N/$^{15}$N, and $^{12}$C/$^{13}$C; and detected CH$_4$, C$_2$H$_6$, C$_2$H$_2$, C$_2$N$_2$, and CO$_2$ as they evaporated from the surface directly below the probe \cite{Niemann2005,Niemann2010}.

\subsection{The Rosetta Orbiter Spectrometer for Ion and Neutral Analysis on Rosetta}
\label{sec:ROSINA}
ROSINA, the Rosetta Orbiter Spectrometer for Ion and Neutral Analysis onboard Rosetta, measured the elemental, isotopic and molecular composition of 67P/Churyumov-Gerasimenko's atmosphere and ionosphere for a good two years, from August 2014 to September 2016 \cite{Balsiger2007}. The ROSINA experiment contained two mass spectrometers, the Double Focusing Mass Spectrometer (DFMS) and the Reflectron Time Of Flight mass spectrometer (RTOF). Although ROSINA was not built for sampling dense atmospheres, it is discussed here because it presents the current state-of-art in high mass resolution in a space application. DFMS was designed to exhibit an extremely high mass resolution $m/\Delta m$ of over 9,000 at 50\% peak height, covering a mass range from $m/z=$12 to 150 with a high dynamic range of 10$^{10}$ and a sensitivity of 10$^{-5}$~A/mbar. RTOF, on the other hand, had a lower mass resolution of $m/\Delta m \sim$5,000 \cite{Scherer2006}, but it covered species from $m/z=$1 up to 1,000 with a dynamic range of 10$^8$ and a sensitivity of 10$^{-4}$~A/mbar \cite{Scherer2006}. DFMS and RTOF measured 63$\times$63$\times$26~cm$^3$ and  114$\times$38$\times$24~cm$^3$, respectively, DFMS weighed 16.2~kg while RTOF weighed 14.7~kg, and DFMS required 19~W whereas RTOF required 24~W of power. It took DFMS 20~s for the measurement of one mass line, resulting in a scan cycle period of $\sim$40~min for an entire mass spectrum. RTOF, which did not step through the different mass lines but recorded them quasi-simultaneously, required 200~s of measuring time for a complete mass spectrum. Ionization was accomplished through electron ionization in both mass spectrometers. For ion detection the DFMS had three independent detectors (a Faraday cup, a multi-channel plate, and an electron multiplier \cite{Berthelier2002}), whereas RTOF contained two fast micro-channel plates detectors \cite{Schletti2001}.

The two mass spectrometers were designed to complement each other with high mass resolution (DFMS) and high time resolution (RTOF). The main difference in operating principle was that whereas DFMS stepped through a mass spectrum with a high mass resolution but a limited mass range, RTOF recorded a complete mass spectrum at once, with an extended mass range but a comparably limited mass resolution. Both spectrometers could be run concurrently and both instruments had a low-resolution and a high-resolution mode. In addition, both spectrometers could measure ions and neutrals, but while DFMS could only operate in one mode at a time, RTOF was built to measure ions and neutrals quasi-simultaneously. Both instruments ran throughout the whole two year mission, with both of them focusing on neutral gas measurements. Since both mass spectrometers were high-resolution mass spectrometers per se, and since they were built for a thin atmosphere which allowed direct measurement of the gaseous species, no additional enhancement techniques (like enrichment cells, noble gas cells, or GC additions) were necessary. Both spectrometers did carry a gas calibration unit, though, which, at certain times, injected a defined quantity of a known gas mixture (He, CO$_2$, and Kr), for parameter optimization and in-flight calibration.

Together DFMS and RTOF measured and characterized a vast amount of different elements, isotopes, and molecules. Over 100 different species were identified, amongst which are over 20 isotopic ratios, several noble gasses, about 50 organic compounds (CH, CHN, CHOS, CHO, CHO$_2$, CHON, CHS, CHS$_2$, CHCl compounds) and about 20 in-organic compounds (e.g., NH$_3$, CH$_4$, H$_2$O, CO, N$_2$, O$_2$, H$_2$S, CO$_2$) \cite{Rubin2019}. For a DFMS high-resolution measurement see for example \cite{Wurz2015}.

\begin{table}
\fontsize{7pt}{7pt}\selectfont
\caption{Details for the mass spectrometers flown on the Galileo probe, the Huygens probe, and on Rosetta. Note that this table presents a simplification of the performance of the instruments.}
\label{tab:GPMS_GCMS_DFMS_RTOF}
\begin{tabular}{lrrrr}
\hline\noalign{\smallskip}
Mission             		& Galileo Probe             & Huygens Probe             & Rosetta                       & Rosetta \\
Instrument          		& GPMS                      & GCMS                      & DFMS                          & RTOF \\
\noalign{\smallskip}\hline\noalign{\smallskip}
Atmospheric entry   		& December 7, 1995          & January 14, 2005          & August 1, 2014                & August 1, 2014 \\ 
MS type:            		& quadrupole                & quadrupole                & magnetic                      & time-of-flight \\
GC addition:        		& no                        & yes                       & no                            & no \\
Noble gas cell:     		& yes                       & yes                       & no                            & no \\
Enrichment cell     		& yes                       & yes                       & no                            & no \\
Ion source:         		& electron ionization       & electron ionization       & electron ionization           & electron ionization \\
Ion detector:       		& electron multiplier       & electron multiplier       & electron multiplier, FC, MCP  & MCP \\ 
Pressure range (external): 	& $\sim$500~mbar--20~bar    & $\sim$3~mbar--$>$1~bar    & $10^{-5}$--$10^{-15}$~mbar	& $10^{-6}$--$10^{-17}$~mbar \\
Pressure range (internal):	& $10^{-4}$--$10^{-13}$~mbar& 							& $10^{-5}$--$10^{-15}$~mbar    & $10^{-6}$--$10^{-17}$~mbar \\
Measurement range   		&                           & 146~km--surface           & $<$1,000~km--surface          & $<$1,000~km--few km \\
Mass range:         		& $m/z=$2--150              & $m/z=$2--141              & $m/z=$12--150                 & $m/z=$1--$\sim$1,000 \\
Mass resolution:    		& unit                      & unit                      & $>$9,000                      & $\sim$5,000 \\
Sensitivity:        		& $\sim10^{12}$--$10^{13}$~counts/s/mbar   & $10^{14}$~counts/s/mbar   & $\sim10^{14}$~counts/s/mbar   & $\sim10^{15}$~counts/s/mbar \\
Dynamic range:      		& $<10^8$                   & $<10^8$                   & $10^{10}$                     & $10^6$/$10^8$ \\
Power:              		& 13~W+12~W                 & 41~W                      & 19~W                          & 24~W \\
Weight:             		& 13.2~kg                   & 17.3~kg                   & 16.2~kg                       & 14.7~kg \\
Size:               		& 			                & 198~mm$\times$470~mm       & 63$\times$63$\times$26~cm$^3$     & 114$\times$38$\times$24~cm$^3$ \\
\noalign{\smallskip}\hline
\end{tabular}
\end{table}

\section{Mass Spectrometry in Space --- State-of-Art}
\label{sec:State_of_Art}
Mass spectrometers can be coarsely divided into low-resolution mass spectrometers and high-resolution mass spectrometers, irrespective of the mass spectrometer type itself. Where one classification ends and the other one begins depends on the isobaric interferences that need to be separated, though, and thus depends on the scientific scope of the mission, which varies from case to case. In any case, low-resolution mass spectrometers are defined as being unable to resolve the necessary isobaric interferences themselves, thus requiring additional sample gas separation systems. As a rule of thumb, for low-resolution mass spectrometers one asks for a mass resolution matching the mass range, e.g., $m/\Delta m \geq m_{\rm max}$, to have unit mass resolution at the highest observed mass.  In contrast, high-resolution mass spectrometers can, ideally,  separate all necessary species even in the presences of isobaric interferences without the need of additional sub-systems. To distinguish isobaric interferences such as $^3$He/HD, CO/N$_2$, $^{20}$Ne/$^{40}$Ar$^{++}$, or $^{22}$Ne/$^{44}$CO$_2$$^{++}$, mass resolutions of 1,000s to 10,000s are required, whereas mass resolutions on the order of 100,000 and more are necessary to resolve species such as C$_x$H$_y$ or even C$_y$H$_y$N$_z$, for example when investigating Titan's atmosphere. In the first case, the class boundary between low- and high-resolution instruments thus lies at a mass resolution of 1,000--10,000, whereas for the second case the class boundary lies beyond 100,000.

Up until about 15 years ago, only low-resolution mass spectrometers were flown on space missions while high-resolution mass spectrometers were confined to laboratory applications. The reason for this was that high-resolution mass spectrometers were still too big, heavy and power-consumptive and operationally too involved to be suitable for placement on a spacecraft. Low resolution mass spectrometers are, though, as mentioned unable to avoid isobaric interferences in the mass spectra. They thus have to be coupled with additional subsystems, e.g., chemical pre-separation. 

Today's available pre-separation techniques work well, but they usually only target a limited range of species. Accordingly, one has to make a pre-selection of the species to be investigated in detail. This is possibly a limitation when an object with little previous knowledge is explored. In addition, while low-resolution mass spectrometers are relatively simple in design, advanced pre-processing techniques (e.g., gas chromatographic systems with several columns) are rather complex systems, both to build and to operate, and they exhibit very high mass and power requirements (see for example the QMS on SAM with a mass of 40~kg and power up to 200~W \cite{Mahaffy2012}). With the technological progress accomplished in the last few decades, high-resolution mass spectrometers have started to overcome many of their technological challenges for space flight. High resolution mass spectrometers have thus started to be considered, proposed, and used in space research. In the following two subsections we present the current state-of-art in mass spectrometry pertaining to atmospheric research, divided into two subsections, which are high-resolution mass spectrometers and low-resolution mass spectrometers, with the latter subsection also containing chemical pre-separation techniques. 

\subsection{High-Resolution Mass Spectrometers}
\label{sec:High_Resolution}
ROSINA with its double-focusing magnetic mass spectrometer DFMS and its reflectron time-of-flight mass spectrometer RTOF was probably the first experiment with high-resolution mass spectrometry flown in space (cf. Section~4.3). DFMS made use of an extensive heritage from high-performance laboratory instruments, rapid technological progress, and modern methods for calculating ion optical properties. DFMS boasted a mass resolution $m/\Delta m$ of over 9,000, allowing for the determination of isotope ratios at the 1\% level \cite{Haessig2017,Rubin2018,Schroeder2019a,Schroeder2019b}. RTOF, with its two ion optical mirrors \cite{Scherer2006,Hohl1999}, belongs to a novel category of time-of-flight mass spectrometers, reflectron time-of-flight mass spectrometers, discussed below. Finally, the CosmOrbitrap development, using an orbitrap mass spectrometer, introduced as a prototype for a space flight \cite{Briois2016}, is presented in the next but one paragraph.

\subsubsection{Time-of-Flight Mass Spectrometers}
\label{sec:Multi_Bounce}
Ions traveling through the drift-tube of a time-of-flight mass spectrometer are separated according to their $m$/$z$ values and arrive at the detector at distinct but predictable times. Ideally, ions with the same $m$/$z$ values would arrive at the exact same time at the detector. In reality, though, ions experience an arrival-time dispersion, because they are not perfectly mono-energetic, and because they contain variances in time, position and velocity (direction and magnitude) when they are ionized and extracted from the ion source towards the drift tube. This results in a broadening of the peak-width in the mass spectrum and thus a reduced mass resolution. As a countermeasure reflectron time-of-flight mass spectrometers were developed \cite{Mamyrin1973}. A reflectron is an ion-optical element, an ion mirror, that uses specially designed electrostatic fields to reflect the incoming ion beam back towards the detector with the goal of focusing ion packets in space and time. In the ion mirror, ions with higher energies penetrate deeper into the retarding electric field, and accordingly spend more time there, than ions with lower energy. This energy-dependent time-delay counterbalances the fact that higher energetic ions pass through the drift-tube faster than lower energetic ions, resulting in a reduced arrival-time dispersion, a reduced peak-width and an improved mass resolution. In addition, by traversing the field-free drift-tube twice, ions of different $m$/$z$ values are dispersed more strongly in time, resulting in an additionally increased mass resolution. Reflectron time-of-flight mass spectrometers can achieve mass resolutions of a few thousand. The  mass spectrometer RTOF on Rosetta was a triple-pass ion-optical system, which used two ion mirrors to fold the ion path three times \cite{Scherer2006}.

Even higher mass resolution can be achieved with multi-bounce mass spectrometers where the ion trajectories are folded several, perhaps many, times within the same ion optical system. This allows for long time-of-flight paths, and thus for high mass resolutions. An example of such a multi-bounce instrument, MASPEX, is currently being built for the Europa Clipper mission of NASA \cite{Brockwell2016}. A flight version of the MASPEX instrument is under development. A form, fit, and function Engineering Model is presently being tested at Southwest Research Institute in San Antonio, Texas, USA. Mass resolutions of $>$30,000 (10\% peak valley definition) can be achieved with the MASPEX instrument. Because lighter $m$/$z$ ion packets can overtake heavier ones, one has to limit the mass window for the analysis the more bounces one exercises, though, i.e., the higher the mass resolution the narrower the usable mass window. For the MASPEX instrument operated at 10 bounces (corresponding to $m$/$\Delta m \approx$~4,500), the usable mass window is about $\pm$14\% of the selected mass, when operated at 26 bounces (corresponding to $m$/$\Delta m \approx$~13,500), the usable mass window is about $\pm$5.5\%.


An alternative to ion-mirrors for time- and space-focusing are electric sensor fields, and \cite{Okumura2004} presented a multi-pass time-of-flight instrument which exhibits a mass resolution of $m/\Delta m <$ 350,000 \cite{Toyoda2003}. Increased mass resolution does not come without some additional costs, though. Multi-pass mass spectrometers are not simple but rather very complex systems to build and to operate. For the outstanding mass resolution of $m/\Delta m$ = 350,000, for example, the MULti-TUrn time-of-flight Mass spectrometer (MULTUM) included no less than 28 carefully tuned electric quadrupole lenses, and up to 500 passes through the ion optical system were required \cite{Shimma2012,Toyoda2003}.

\subsubsection{Orbitrap-Based Mass Spectrometers}
\label{sec:Orbitraps}
Orbitrap mass spectrometers are of a non-scanning nature, come in a compact design, have a good detection dynamic, and exhibit extremely high mass resolution and precision. \cite{Horst2012}, for example, presented a laboratory orbitrap mass spectrometer with a mass resolutions of $m/\Delta m$ = 100,000 for up to $m/z=$400 that was used to analyze chemical processes simulating Titan's atmosphere. \cite{Briois2016} presented a prototype instrument for the JUICE mission of ESA, an orbitrap mass spectrometer optimized for {\it in situ} analysis of dust from airless bodies in the Solar System, with a mass resolution of $m/\Delta m$ = 474,000 at $m/z=$9 reducing to $m/\Delta m$ = 60,000 at $m/z=$400, following the theoretical trend of the mass resolution of orbitrap systems of $m/\Delta m \propto m^{-1/2}$ \cite{Makarov2000}. A concept for a combination of a linear ion trap and an orbitrap mass spectrometer has been proposed for measurements of thin planetary atmospheres \cite{Arevalo2016a}. 


\subsection{Low-Resolution Mass Spectrometers}
\label{sec:Low_Resolution}
Low resolution mass spectrometers are defined as having sufficient mass resolution to separate all adjacent mass lines in the covered mass range, but not to resolve isobaric interferences that may occur in the sample. They offer an alternative approach to high-resolution mass spectrometers, but they should be coupled to a more elaborate sample preparation using pre-separation, pre-selection, cleaning, or enrichment  systems to achieve a comparable performance concerning resolution, and species identification.  For the sample preparation several options exists, which have been discussed in Section~\ref{sec:Additional}. The main task of the sample preparation system is to enable the isolation of species or groups of them, to avoid isobaric interferences (because mass peaks with almost identical $m$/$z$ values cannot be resolved), to provide a diagnostic measure of composition, and to extend the total dynamic range. As for the mass spectrometer itself, any mass spectrometer from the categories presented in Section~\ref{Mass_Analyser} can be used. 

\subsection{Gas Chromatograph Mass Spectrometers}
\label{sec:Huygens}
Gas chromatography coupled with mass spectrometry (GCMS) is a very powerful analysis method, and  has been used on planetary science payloads such as Viking \cite{Rushneck1978}, Cassini-Huygens \cite{Niemann2002}, and Sample Analysis at Mars (SAM) of the Mars Science Laboratory \cite{Mahaffy2012} on Curiosity. 

The Cassini-Huygens mission consisted of two main elements: The Cassini orbiter, which orbited Saturn for over 13 years (June 2004 -- September 2017), and the Huygens probe, which descended through to Titan's atmosphere all the way to the surface  in January 2005. The lander probe carried a Gas Chromatograph Mass Spectrometer (GCMS) experiment designed to identify and characterize the chemical composition of the atmosphere of Titan and to determine the isotope ratios of the major gaseous constituents \cite{Niemann2002}. The instrument measured atmospheric species for 148~minutes during descent (from $\sim$146~km down to the surface). Having survived the impact on the surface, GCMS measured species evaporated from the surface  for an additional 72~minutes until contact was lost with the Cassini orbiter, which relayed the communications \cite{Lebreton2005}. The GCMS was of cylindrical shape, with a diameter of 198~mm and a height of 470~mm, weighed 17.3~kg, and required on average 41~W of power. The mass analyzer was capable of measuring ions from 2 to 141~u/e with unit mass resolution and a mixing ratio detection limit on the order of 10$^{-8}$. Stepping was performed at unit mass steps with a single mass step taking 5~ms to acquire, resulting in a mass scan cycle of 937.5~ms. In-flight, the mass scan cycle doubled to 1.875~s due to transmission constraints. Ionization was achieved through electron impact ionization and ions were detected by a secondary electron multiplier detection system.

\subsection{Tunable Laser Spectrometer}
\label{sec:Tunable_Laser_Spectrometer}
Tunable Laser Spectrometers (TLS) \cite{Durry2002} pose a well-suited complement to a mass spectrometer system by measuring the isotopic ratios of some selected molecules, e.g. H$_2$O, NH$_3$, CH$_4$, PH$_3$, CO, CO$_2$, SO$_2$, and OCS, to a high accuracy. TLS measure the absorption of light of a  laser in a cavity at specific wavelengths in the near infra-red (IR) to mid-IR spectral region in a direct, non-invasive, unambiguous manner. Targeted species can be detected by scanning over their rovibrational spectral lines without the concern of interferences. In this way, TLS reach ultra-high spectral resolutions (on the order of 0.0001 cm$^{-1}$), high accuracies of a few \% for species abundances and about one per mille for isotope determinations, and remarkable sensitivities at the sub-ppb level for gas detection, while still being small in mass and volume. The main isotopic ratios targeted by TLS systems used in atmospheric analyses are D/H, $^{18}$O/$^{16}$O, $^{17}$O/$^{16}$O, $^{13}$C/$^{12}$C, $^{15}$N/$^{14}$N, and $^{34}$S/$^{32}$S. Exemplary missions which carried TLS systems to space are the ExoMars mission, where a TLS was used to analyze the Martian atmosphere \cite{LeBarbu2004}, the Phobos Grunt mission, which was designed to examine the martian moon Phobos \cite{Durry2010}, and the Mars Science Laboratory mission, a TLS system on which measured the isotopic ratios of D/H and of $^{18}$O/$^{16}$O in water, and of $^{13}$C/$^{12}$C, $^{18}$O/$^{16}$O, $^{17}$O/$^{16}$O, and $^{13}$C$^{18}$O/$^{12}$C$^{16}$O in carbon dioxide in the Martian atmosphere \cite{Webster2013}.

\subsection{Helium Abundance Detector}
\label{sec:Helium_Abundance_Detector}
A Helium Abundance Detector (HAD), another well-suited supplement to a mass spectrometer system, is an optical interferometer used to determine the helium abundance in a given sample. This is accomplished through a two-beam interferometer, where one beam passes through a reference gas mixture of known composition, and  where the other beam passes through an atmospheric gas sample to be analyzed. The difference in the optical path gives the difference in refractive index between the reference gas and the atmospheric gas. On Galileo, a HAD was used for the precise (0.1\%) determination of the helium abundance in the Jovian atmosphere from 2 to 10~bars \cite{vonZahn1992}. The measurement relied on the fact that the Jovian atmosphere consists mostly of H$_2$ and He (to more than 99.5\%), and that the refractive index thus poses a direct measure of the He/H$_2$ ratio. The instrument was very compact, weighing only 1.4~kg and requiring power of less than 1~W. The Galileo HAD had an expected He/H$_2$ ratio accuracy of $\pm$0.0015. The measurement yielded a  He mole fraction of 0.1350$\pm$0.0027, with a thus somewhat lower accuracy than expected, but still better than was possible by a mass spectrometric measurement at the time \cite{vonZahn1998}.

\section{Possible Descent Probe Implementation}
\label{sec:Recommendation}
Measuring the ice giants' chemical composition gives valuable insight into the gas and the dust reservoir available and the processes at work when our Solar System formed. In particular, knowledge of the ice giants' gross atmospheric composition and atmospheric isotopic ratios (both of which are representative of their bulk counterparts) can help us distinguish between currently competing formation scenarios, e.g., \cite{Mizuno1980,Pollack1996,Boss1997,Boss2001}. Species  of interest in this context are H$_2$, HD and He, including their abundance and isotopic ratios, the in-organic compounds CH$_4$, NH$_3$, H$_2$S, PH$_3$, H$_2$O, CO, CO$_2$, AsH$_3$, GeH$_4$, the organic compounds C$_2$H$_2$ and C$_2$H$_6$, the isotopic ratios of $^{13}$C/$^{12}$C, $^{15}$N/$^{14}$N, $^{16}$O/$^{17}$O, $^{18}$O/$^{16}$O, and $^3$He/$^4$He, and finally the heavy noble gases Ne, Ar, Kr and Xe, including their abundant isotopic ratios.

A descent probe offers the possibility to determine the chemical abundance and isotopic ratios of these species by {\it in situ} sampling of the atmospheric gas. Ideally, a full mass spectrum is recorded at altitude steps with sufficiently fine stepping. An instrument designed to measure these species satisfying the above-stated scientific requirements requires high throughput (due to the limited measurement time a descent probe offers), high sensitivity and dynamic range (due to the extremely low mixing-ratio of some of these species), and a sufficient mass resolution and accuracy (due to the isobaric interferences of some of these species). Resources, in particular power, mass, and presumably data rate are likely to be severely restricted for a Uranus/Neptune atmospheric probe, even more than they were for the Galileo and Cassini probes, due to the ice giants' distant location. 

\begin{figure}
  \includegraphics[width=1.00\textwidth]{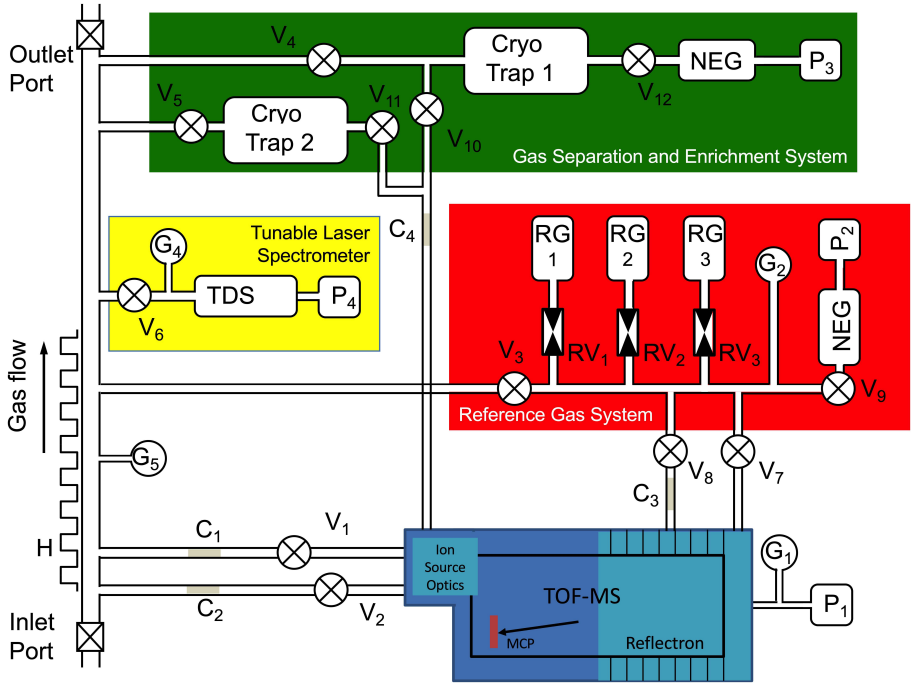}
  \caption{Proposed descent probe mass spectrometer experiment, with four major functional units indicated by color: the time-of-flight mass spectrometer (TOF-MS; blue), the gas separation and enrichment system (green), the reference gas system (red), and the tunable laser spectrometer (yellow). The abbreviations are as follows: valves (V), regulating valves (RV), pressure gauges (G), heater (H), conductance limiters (C), pumps (P), gas reservoirs (RG), and non-evaporative getter (NEG).}
  \label{fig:System}
  \end{figure}

To realize the scientific objectives of a Uranus/Neptune atmospheric probe within realistic resources on the probe we consider a system comprised out of four subsystems as shown in Figure~\ref{fig:System}: i) a gas separation and enrichment system, ii) a reference gas system, iii) a mass spectrometer, and iv) a tunable laser spectrometer. In addition, careful attention to the gas handling system has to be made. The atmospheric gas should pass through the system easily, should not condense anywhere, causing delayed measurements or history effects. The gas should not pile up at places, necessitating good pumping inside the instrument. Moreover, it has to be expected that a large fraction of the descent is through various cloud layers \cite{Atreya2020}, and a fraction of the volatiles will be in the condensed phase. Therefore, the gas inlet system has to be heated enough to volatilize every cloud particle in the gas inlet. Ideally, the gas inlet has two entrance channels, one that allows for the entry of gas and condensed particles at the same time, and one allowing only for gas entry. Be switching between these two channels in regular intervals one obtains the composition of the volatiles and of the clouds in the part of the atmosphere covered by the descent trajectory. Recent laboratory measurements have shown that noble gases condense well on graphene surfaces kept at cryogenic temperatures similar to the ones expected during the descent \cite{Maiga2018}. Thus it might be that the cloud particles present a similar storage of noble gases in Uranus/Neptune's atmosphere, which can be investigated with a two channel inlet system. Finally, the heterogeneity of composition observed in Jupiter's atmosphere for ammonia \cite{Bolton2017} might just be variability of the fraction of volatile species being in the gas phase or in the condensed phase. By measuring both phases, we would be able to resolve this issue, and arrive at measurements closer to the deep atmosphere composition. 

The mass spectrometer is a low-resolution time-of-flight mass spectrometer, with a mass resolution $m/\Delta m$ of about 1,000 (sufficient to separate $^3$He from HD) and a fast acquisition rate, similar to the mass spectrometer designed for the Luna Resurs mission \cite{Wurz2012}, which would fit into the available resources and fulfills the stated science requirements \cite{Hofstadter2017}. Because this instrument is a TOF-MS no scanning of the mass spectrum is necessary, resulting in an improved sensitivity by a factor of 100--1000 compared to the Galileo Probe mass spectrometer. To establish the required accuracies for chemical and, in particular, isotopic composition, the mass spectrometer is coupled with valves to a reference gas system containing three reservoirs of calibration gases of accurately known chemical and isotopic composition. These gases will be used to calibrate the mass spectrometer shortly before entry into the atmosphere when the actual activities are commenced. 

In addition, the mass spectrometer is coupled to a gas separation and enrichment system that contains two cryotraps \cite{Brockwell2016}. During nominal measurement proceedings the cryotraps will collect and purify noble gases, which are at a later point released to the mass spectrometer for analysis, three times for cryotrap 1 and two times for cryotrap 2. The noble gas enhancement is achieved by using a combination of cryotrap 1, ion pump P3, and a non-evaporable getter (NEG: SAES 172). The NEG removes all constituents except methane and the noble gases. The cryotrap 1 traps the products of the NEG process, except for helium and some neon. The ion pump then operates to pump away the helium, which is the second highest source of gas, thus enhancing the remaining noble gases ~200 times. Helium and neon are measured using a separate mode. Cryotrap 2 works as an enhancement cell for most gases other than H$_2$ and He. Use of a GC system for the selective measurement of noble gases is not considered because the cryotrap system will allow for enhancements of heavy noble gases by factor of about 5000, whereas the GC system would allow for a factor of 100 \cite{Hofer2015} and will need more power to operate. 


Part of the proposed mass spectrometer experiment is a tunable laser mass spectrometer that can sample the gas directly to investigate selected isotope ratios, e.g. D/H, $^{13}$C/$^{12}$C, $^{18}$O/$^{16}$O, and $^{17}$O/$^{16}$O, depending on the selected laser system. The four subsystems comprising the complete mass spectrometer experiment are combined by a complex vacuum system with a primary inlet and a primary outlet port as well as two conductance passages to cover the whole pressure range during the atmospheric descent ($\sim$1$\cdot$10$^{-4}$~mbar to $\sim$20~bar). If resources allow, a Helium Abundance Detector should be added, to provide redundancy to the H$_2$/He measurement because of the high importance of this measurement. 

Table~\ref{tab:Operation_sequence} presents a possible operation sequence for a mass spectrometer experiment on an atmospheric descent probe considering a descent time of about 60 minutes. Note that this sequence was designed for an atmospheric probe entering Saturn's atmosphere, but a similar operation sequence is expected to be applicable to an Uranus or Neptune descent probe. Accumulation times for mass spectra vary between 5 and 10~seconds, i.e., a full mass spectrum will be recorded every 5 to 10~seconds. Based on the descent speed, vertical resolutions for consecutive mass spectra vary between a few 100~m and 25~km (see Table~\ref{tab:Operation_sequence}).

The operations of the mass spectrometer experiment during descent through the atmosphere are divided into six phases, and a pre-phase at the atmospheric entry: 
\begin{itemize}
	\item Phase 0: comprises the pressure range where `empty space' turns into the planetary atmosphere. 
	\item Phase 1--2: High-speed descent through the upper atmosphere
	\item Phase 3--5: Low-speed descent through the atmosphere proper
	\item Phase 6: Final descent until loss of contact
\end{itemize}

{\em Phase 0:} During this phase around 14 mass spectra will be recorded in a region of the atmosphere, from 1500~km down to 450~km, where  photon-induced chemical reactions predominately take place, with a complete mass spectrum being available every 30~seconds. At the end of Phase 0 the cryotrap 2 will be read out with the enhanced signal of hydrocarbons from the upper atmosphere.  

{\em Phases 1--2:} Regular atmospheric measurements  start at a pressure of about 1$\cdot$10$^{-4}$~mbar. In the beginning, when the atmospheric probe is still in free-fall and when the atmospheric pressure is still very low, the integration time is optimized to accumulate as much rarefied gas as possible while still providing a decent vertical resolution. Approximately 11 mass spectra will be recorded in the Phase 1  part of the atmosphere, the traversing of which lasts a few minutes. At about 400~mbar, the drogue parachute is opened commencing a short Phase~2,  with only 1 mass spectrum being directly recorded. At the end of Phase~2 the cryotrap 2 will be read out a second time, again with the enhanced signal of hydrocarbons from this part of the atmosphere.  During these phases, all measurements are performed in low sensitivity mode, putting emphasis on the accurate measurement of the H$_2$/He ratio in the atmosphere. 

{\em Phases 3--5:} As soon as the drogue parachute has fulfilled its mission, the main parachute is deployed and high vertical resolution measurements begin. As the descent probe was slowed down by the main parachute, the operations switch from low vertical resolution to high vertical resolution mode, where the integration time is optimized and the vertical resolution corresponds a few 100~m.  In this part of the atmosphere the probe collects approximately 141 mass spectra in total. At a pressure of a little over 1~bar the probe is expected to enter the cloud layer \cite{Atreya2019,Mousis2019}. The vertical resolution is maximized at this location. Moreover, the main gas sampling pipe is equipped with a dedicated heater to assure that all condensed particles will completely vaporize. Phase~5 measurements are planned to start at the expected pressure level of the cloud layer. During these phases, the instrument is operated in high sensitivity mode, to put emphasis on the less abundant species of the  atmosphere. This mode is accomplished by blanking out the low-mass part of the mass spectrum (i.e., reducing H$_2$ and He by a factor 1000 in the mass spectrum) and increasing the detector gain. At the end of Phase~3, 4, and 5 there is always the read-out of cryotrap 1, to get highly sensitive and very accurate measurements of the noble gases and their isotopes at three pressure levels in the atmosphere. 

{\em Phase 6:} This is the final phase where the probe descends below the nominal pressure level. The mass spectrometer accumulates as many mass spectra as possible before communications with the mother spacecraft are no longer possible, because the spacecraft has moved over the horizon seen by the descent probe. As mentioned above, this is expected to happen about 60~minutes after measurements have initially begun. In this final, high cadence, mode, the probe is expected to be able to record almost 30 mass spectra resulting in a grand total of about 250 complete atmospheric mass spectra for the complete descent.

The estimated accuracies to be obtained for the operations described above are given in Table~\ref{tab:Accuracies}, which shows that the stated scientific requirements for an atmospheric probe, see \cite{Mousis2014,Atreya2018,Atreya2020,Cavalie2020,Simon2020}, can be met by such a system. In addition to the mass spectrometric measurements, and independently from it, the Tunable Laser Spectrometer and the Helium Abundance Detector will sample the same atmospheric gas directly.

\begin{table}
\caption{Sample operation sequence of mass spectrometer experiment  during the atmospheric descent. Pressure is given at the beginning of a phase. }
\label{tab:Operation_sequence}
\begin{tabular}{lrrrrrr}
\hline\noalign{\smallskip}
Phase & Altitude  &	Time span, from--to [sec] & Pressure & Integration  & Vertical & Number of  \\
           &   [km]     &	                 &   [mbar] &   time [sec] &  resolution [km] & mass spectra \\
\noalign{\smallskip}\hline\noalign{\smallskip}
0	&	1500	&	--414.18 -- 0	&	1.00$\cdot 10^{-7}$	&	30	&			&	14 	\\
1	&	450		&	0 -- 172		&	1.00$\cdot 10^{-4}$	&	15	&	76.05	&	11 	\\
2	&	15		&	172 -- 183	&	4.00$\cdot 10^{2}$	&	12	&	38.02	&	1	\\
3	&	14		&	183 -- 235		&	4.20$\cdot 10^{2}$			&	11	&	0.21	&	5	\\
4	&	10		&	235 -- 581	&	5.00$\cdot 10^{2}$	&	10	&	0.12	&	35	\\
5	&	--13		&	581 -- 2101	&	1.60$\cdot 10^{3}$		&	15  	&	0.23	&	101	\\
6	&	--140	&	2101 -- 3684	&	2.40$\cdot 10^{4}$	&	20	&	1.09	&	29	\\
\noalign{\smallskip}\hline
\end{tabular}
\end{table}


\section{Conclusions}
\label{sec:Conclusions}
In this paper we reviewed  mass spectrometry experiments as they pertain to an atmospheric probe for Uranus and Neptune. Moreover, we reviewed current mass spectrometry developments of neutral gas mass spectrometers for space instrumentation, again for possible use on an atmospheric probe. Missions to Uranus and Neptune pose severe limitations in resources available to instrumentation, much more so than for the Galileo probe at Jupiter and the Huygens probe at Titan, which restricts the choices for instrumentation. We proposed a possible mass spectrometer experiment that would fulfill all the stated science requirements, and is expected to provide even better results than was possible for previous atmospheric probes, which is based on technology of the late 1980-ties.   


\begin{acknowledgements}
A. Vorburger and P. Wurz gratefully acknowledge the financial support by the Swiss National Science Foundation.
\end{acknowledgements}

\section*{Conflict of interest}
The authors declare that they have no conflict of interest.

\section*{Appendix}

\begin{table}
\caption{Accuracies for selected species measured during the different phases described in the main text, for the mass spectrometry experiment shown Figure~\ref{tab:Operation_sequence}.}
\label{tab:Accuracies}
\begin{adjustbox}{angle=0}
\begin{tabular}{cccccccccc}
\hline\noalign{\smallskip}
Phase 1	&	Accuracy		& Isotopes & Accuracy\\
\noalign{\smallskip}\hline\noalign{\smallskip}
H$_2$, low sensitivity	&	5.00\%	& HD/H$_2$		& 0.90\%			\\
He, low sensitivity	&	5.00\%		&	$^3$He/$^4$He	& 1.70\% \\
He/H$_2$	&	1.50\%		\\
CH$_4$, low sensitivity	&	5.00\%	&	$^{12}$C/$^{13}$C &	1.90\%	\\
H$_2$S, low sensitivity	&	5.10\%		\\
C$_2$H$_2$, low sensitivity	&	9.00\%		\\
C$_2$H$_6$, low sensitivity	&	8.60\%		\\
Ne, low sensitivity	&	5.70\%		\\
Ar, low sensitivity	&	5.00\%			\\
Kr, low sensitivity	&	12.10\%		\\
Xe, low sensitivity	&	41.80\%		\\
	
\hline\noalign{\smallskip}
Phase 2	&	Accuracy		& Isotopes & Accuracy \\
\noalign{\smallskip}\hline\noalign{\smallskip}
H$_2$, low sensitivity	&	5.00\%			\\
HD/H$_2$	&	0.90\%		\\
He, low sensitivity	&	5.00\%	&	$^3$He/$^4$He & 1.70\%			\\
He/H$_2$	&	1.50\%		\\
CH$_4$, low sensitivity	&	5.00\%	&	$^{12}$C/$^{13}$C & 1.90\%	 \\
NH$_3$, low sensitivity	&	5.00\%	&	$^{14}$N/$^{15}$N &	12.60\% 	  \\
H$_2$S, low sensitivity	&	5.10\%		\\
C$_2$H$_2$, low sensitivity	&	9.00\%		\\
C$_2$H$_6$, low sensitivity	&	8.60\%		\\
Ne, low sensitivity	&	5.70\%		\\
Ar, low sensitivity	&	5.00\%		\\
Kr, low sensitivity	&	12.10\%		\\
Xe, low sensitivity	&	41.80\%		\\

\hline\noalign{\smallskip}
Phase 3 / 4 / 5	&	Accuracy		& Isotopes & Accuracy \\
\noalign{\smallskip}\hline\noalign{\smallskip}
CH$_4$, high sensitivity	&	5.00\%	&	$^{12}$C/$^{13}$C & 0.30\% \\
NH$_3$, high sensitivity	&	5.00\%		&	$^{14}$N/$^{15}$N  &	1.50\%  \\
H$_2$O at 2~bar, high sensitivity	&	6.50\%		\\
H$_2$S, high sensitivity	&	5.00\%		\\
CO, high sensitivity	&	163.80\%		\\
CO$_2$, high sensitivity	&	457.70\%		\\
PH$_3$ at $\sim$1~bar, high sensitivity	&	5.10\%		\\
AsH$_3$,  high sensitivity	&	69.20\%			\\
GeH$_4$, high sensitivity	&	445.00\%		\\
C$_2$H$_2$, high sensitivity	&	5.10\%		\\
C$2$H$6$, high sensitivity	&	5.10\%		\\
Ne, high sensitivity	&	5.00\%		\\
Ar, high sensitivity	&	5.00\%		\\
Kr, high sensitivity	&	5.20\%		\\
Xe, high sensitivity	&	6.10\%		\\

\hline\noalign{\smallskip}
Phase 6	&	Accuracy	&	Isotopes	&	Accuracy	\\
\noalign{\smallskip}\hline\noalign{\smallskip}
CH$_4$, high sensitivity	&	5.00\%		&	$^{12}$C/$^{13}$C	&	0.20\%	\\
NH$_3$, high sensitivity	&	5.00\%		&	$^{14}$N/$^{15}$N	&	1.04\%	\\
H$_2$O at 10~bar, high sensitivity	&	5.00\%	&	$^{16}$O/$^{17}$O	&	7.40\%	\\
           &		&	$^{16}$O/$^{18}$O		&	3.10\%	\\
H$_2$S, high sensitivity	&	5.00\%		\\
CO, high sensitivity	&	110.50\%		\\
CO$_2$, high sensitivity	&	308.60\%		\\
PH$_3$, high sensitivity	&	5.00\%	\\
AsH$_3$,  high sensitivity	&	46.80\%	\\
GeH$_4$, high sensitivity	&	300.00\%		\\
C$_2$H$_2$, high sensitivity	&	5.00\%		\\
C$_2$H$_6$, high sensitivity	&	5.00\%		\\
Ne, high sensitivity	&	5.00\%		&	$^{20}$Ne/$^{21}$Ne	&	5.40\%	\\
&		&	$^{21}$Ne/$^{21}$Ne	&	5.40\%		\\
Ar, high sensitivity	&	5.00\%			\\
Kr, high sensitivity	&	5.10\%	\\
Xe, high sensitivity	&	5.50\%		\\
\end{tabular}
\end{adjustbox}
\end{table}
\begin{table}
\caption{Continued...}
\label{tab:Accuracies}
\begin{adjustbox}{angle=0}
\begin{tabular}{cccccccccc}
\hline\noalign{\smallskip}
Noble gases (3 times cryotrap)	&	Accuracy	\\
\noalign{\smallskip}\hline\noalign{\smallskip}
Ar, enriched	&		\\
$^{36}$Ar/$^{38}$Ar	&	0.10\%	\\
Kr, enriched	&		\\
$^{78}$Kr/Kr$_{\mathrm{tot}}$	&	1.06\%	\\
$^{80}$Kr/Kr$_{\mathrm{tot}}$	&	1.13\%	\\
$^{82}$Kr/Kr$_{\mathrm{tot}}$	&	0.46\%	\\
$^{83}$Kr/Kr$_{\mathrm{tot}}$	&	0.28\%	\\
$^{84}$Kr/Kr$_{\mathrm{tot}}$	&	0.23\%	\\
$^{86}$Kr/Kr$_{\mathrm{tot}}$	&	0.21\%	\\
Xe, enriched	&		\\
$^{124}$Xe/Xe$_{\mathrm{tot}}$	&	6.18\%	\\
$^{126}$Xe/Xe$_{\mathrm{tot}}$	&	8.74\%	\\
$^{128}$Xe/Xe$_{\mathrm{tot}}$	&	6.28\%	\\
$^{129}$Xe/Xe$_{\mathrm{tot}}$	&	1.19\%	\\
$^{130}$Xe/Xe$_{\mathrm{tot}}$	&	0.85\%	\\
$^{131}$Xe/Xe$_{\mathrm{tot}}$	&	0.87\%	\\
$^{132}$Xe/Xe$_{\mathrm{tot}}$	&	0.49\%	\\
$^{134}$Xe/Xe$_{\mathrm{tot}}$	&	0.60\%	\\
$^{136}$Xe/Xe$_{\mathrm{tot}}$	&	0.74\%	\\
\noalign{\smallskip}\hline
\end{tabular}
\end{adjustbox}
\end{table}

\bibliographystyle{spmpsci}
\bibliography{bibliography}

\begin{thebibliography}{100}
\providecommand{\url}[1]{{#1}}
\providecommand{\urlprefix}{URL }
\expandafter\ifx\csname urlstyle\endcsname\relax
  \providecommand{\doi}[1]{DOI~\discretionary{}{}{}#1}\else
  \providecommand{\doi}{DOI~\discretionary{}{}{}\begingroup
  \urlstyle{rm}\Url}\fi

\bibitem{Arevalo2016a}
{Arevalo} R.~D., J., {Danell}, R.M., {Gundersen}, C., {Hovmand}, L.,
  {Southard}, A.E., {Tan}, F., {Grubisic}, A., {Brinckerhoff}, W.B., {Getty},
  S., {Mahaffy}, P.R., {Cottin}, H., {Briois}, C., {Collin}, F., {Szopa}, C.,
  {Vuitton}, V., {Makarov}, A., {Reinhardt-Szyba}, M.: {Advanced Resolution
  Organic Molecule Analyzer (AROMA): Simulations, Development and Initial
  Testing of a Linear Ion Trap-Orbitrap Instrument for Space}.
\newblock In: 3rd International Workshop on Instrumentation for Planetary
  Missions (2016)

\bibitem{Atreya2018}
Atreya, S.K., Crida, A., Guillot, T., Lunine, J.I., Madhusudhan, N., Mousis,
  O.: {The Origin and Evolution of Saturn, with Exoplanet Perspective}, pp.
  5--43.
\newblock Cambridge Planetary Science. Cambridge University Press (2018).
\newblock \doi{10.1017/9781316227220.002}

\bibitem{Atreya2020}
Atreya, S.K., Hofstadter, M.H., In, J.H., Mousis, O., Reh, K., Wong, M.H.:
  {Deep Atmosphere Composition, Structure, Origin, and Exploration, with
  Particular Focus on Critical in situ Science at the Icy Giants}.
\newblock Space Science Review  (2020).
\newblock This issue

\bibitem{Atreya2019}
Atreya, S.K., Hofstadter, M.H., Reh, K., In, J.H., Mousis, O., Wong, M.H.: {Icy
  giant planet exploration: Are entry probes essential?}
\newblock Acta Astronautica \textbf{162}, 266--274 (2019).
\newblock \doi{https://doi.org/10.1016/j.actaastro.2019.06.020}.
\newblock
  \urlprefix\url{http://www.sciencedirect.com/science/article/pii/S0094576519302668}

\bibitem{Avice2019}
Avice, G., Belousov, A., Farley, K.A., Madzunkov, S., Simcic, J., Nikoli\'{c},
  D., Darrach, M.R., Sotin, C.: {High-precision measurements of krypton and
  xenon isotopes with a new static-mode quadrupole ion trap mass spectrometer}.
\newblock {Journal of Analytical Atomic Spectrometry} \textbf{34}, 104--117
  (2019).
\newblock \doi{10.1039/C8JA00218E}

\bibitem{Balsiger2007}
Balsiger, H., Altwegg, K., Bochsler, P., Eberhardt, P., Fischer, J., Graf, S.,
  J{\"a}ckel, A., Kopp, E., Langer, U., Mildner, M., M{\"u}ller, J., Riesen,
  T., Rubin, M., Scherer S. nd~Wurz, P., W{\"u}thrich, S., Arijs, E., Delanoye,
  S., Keyser, J.D., Neefs, E., Nevejans D.and~R{\`e}me, H., Aoustin, C.,
  Mazelle, C., M{\'e}dale, J.L., Sauvaud, J.A., Berthelier, J.J., Bertaux,
  J.L., Duvet, L., Illiano, J.M., Fuselier, S.A., Ghielmetti, A.G., Magoncelli,
  T., Shelley, E.G., Korth, A., Heerlein, K., Lauche, H., Livi, S., Loose, A.,
  Mall, U., Wilken, B., Gliem, F., Fiethe, B., Gombosi, T.I., Block, B.,
  Carignan, G.R., Fisk, L.A., Waite, J.H., Young, D.T., Wollnik, H.: {Rosina --
  Rosetta Orbiter Spectrometer for Ion and Neutral Analysis}.
\newblock Space Science Reviews \textbf{128}(1), 745--801 (2007).
\newblock \doi{10.1007/s11214-006-8335-3}

\bibitem{Balsiger1971}
{Balsiger}, H., {Eberhardt}, P., {Geiss}, J., {Kopp}, E.: {A Mass Spectrometer
  for the Simultaneous Measurement of the Neutral and the Ion Composition of
  the Upper Atmosphere}.
\newblock Review of Scientific Instruments \textbf{42}(4), 475--476 (1971).
\newblock \doi{10.1063/1.1685134}

\bibitem{Barabash2013}
{Barabash}, S., {Wurz}, P., {Brandt}, P., {Wieser}, M., {Holmstr{\"o}m}, M.,
  {Futaana}, Y., {Stenberg}, G., {Nilsson}, H., {Eriksson}, A., {Tulej}, M.,
  {Vorburger}, A., {Thomas}, N., {Paranicas}, C., {Mitchell}, D.G., {Ho}, G.,
  {Mauk}, B.H., {Haggerty}, D., {Westlake}, J.H., {Fr{\"a}nz}, M., {Krupp}, N.,
  {Roussos}, E., {Kallio}, E., {Schmidt}, W., {Szego}, K., {Szalai}, S.,
  {Khurana}, K., {Jia}, X., {Paty}, C., {Wimmer-Schweingruber}, R.F., {Heber},
  B., {Kazushi}, A., {Grande}, M., {Lammer}, H., {Zhang}, T., {McKenna-Lawlor},
  S., {Krimigis}, S.M., {Sarris}, T., {Grodent}, D.: {Particle Environment
  Package (PEP)}.
\newblock In: European Planetary Science Congress, pp. EPSC2013--709 (2013)

\bibitem{Bennett1950}
Bennett, W.H.: {Radiofrequency Mass Spectrometer}.
\newblock Journal of Applied Physics \textbf{21}(2), 143--149 (1950).
\newblock \doi{10.1063/1.1699613}

\bibitem{Benninghoven1987}
Benninghoven, A., R{\"u}denauer, F.G., Werner, H.W.: Secondary ion mass
  spectrometry---Basic concepts, instrumental aspects, applications and trends.
\newblock Surface and Interface Analysis. Wiley, New York (1987).
\newblock \doi{10.1002/sia.740100811}.
\newblock
  \urlprefix\url{https://onlinelibrary.wiley.com/doi/abs/10.1002/sia.740100811}

\bibitem{Berthelier2002}
Berthelier, J.J., Illiano, J.M., Nevejans, D., Neefs, E., Arijs, E., Schoon,
  N.: {High resolution focal plane detector for a space-borne magnetic mass
  spectrometer}.
\newblock International Journal of Mass Spectrometry \textbf{215}(1), 89--100
  (2002).
\newblock \doi{https://doi.org/10.1016/S1387-3806(02)00527-4}.
\newblock
  \urlprefix\url{http://www.sciencedirect.com/science/article/pii/S1387380602005274}.
\newblock Detectors and the Measurement of Mass Spectra

\bibitem{Bolton2017}
Bolton, S.J., Adriani, A., Adumitroaie, V., Allison, M., Anderson, J., Atreya,
  S., Bloxham, J., Brown, S., Connerney, J.E.P., DeJong, E., Folkner, W.,
  Gautier, D., Grassi, D., Gulkis, S., Guillot, T., Hansen, C., Hubbard, W.B.,
  Iess, L., Ingersoll, A., Janssen, M., Jorgensen, J., Kaspi, Y., Levin, S.M.,
  Li, C., Lunine, J., Miguel, Y., Mura, A., Orton, G., Owen, T., Ravine, M.,
  Smith, E., Steffes, P., Stone, E., Stevenson, D., Thorne, R., Waite, J.,
  Durante, D., Ebert, R.W., Greathouse, T.K., Hue, V., Parisi, M., Szalay,
  J.R., Wilson, R.: {Jupiter's interior and deep atmosphere: The initial
  pole-to-pole passes with the Juno spacecraft}.
\newblock Science \textbf{356}(6340), 821--825 (2017).
\newblock \doi{10.1126/science.aal2108}

\bibitem{Boss1997}
Boss, A.P.: {Giant planet formation by gravitational instability}.
\newblock Science \textbf{276}(5320), 1836--1839 (1997).
\newblock \urlprefix\url{www.scopus.com}.
\newblock Cited By :529

\bibitem{Boss2001}
Boss, A.P.: {Formation of planetary-mass objects by protostellar collapse and
  fragmentation}.
\newblock Astrophysical Journal \textbf{551}(2 PART 2), L167--L170 (2001).
\newblock \urlprefix\url{www.scopus.com}.
\newblock Cited By :103

\bibitem{Briois2016}
Briois, C., Thissen, R., Thirkell, L., Aradj, K., Bouabdellah, A., Boukrara,
  A., Carrasco, N., Chalumeau, G., Chapelon, O., Colin, F., Coll, P., Cottin,
  H., Engrand, C., Grand, N., Lebreton, J.P., Orthous-Daunay, F.R., Pennanech,
  C., Szopa, C., Vuitton, V., Zapf, P., Makarov, A.: Orbitrap mass analyser for
  in situ characterisation of planetary environments: Performance evaluation of
  a laboratory prototype.
\newblock Planetary and Space Science \textbf{131}, 33 -- 45 (2016).
\newblock \doi{https://doi.org/10.1016/j.pss.2016.06.012}.
\newblock
  \urlprefix\url{http://www.sciencedirect.com/science/article/pii/S0032063316300058}

\bibitem{Brockwell2016}
{Brockwell}, T.G., {Meech}, K.J., {Pickens}, K., {Waite}, J.H., {Miller}, G.,
  {Roberts}, J., {Lunine}, J.I., {Wilson}, P.: {The mass spectrometer for
  planetary exploration (MASPEX)}.
\newblock In: 2016 IEEE Aerospace Conference, pp. 1--17 (2016).
\newblock \doi{10.1109/AERO.2016.7500777}

\bibitem{Buchanan1987}
Buchanan, M.V., Wise, M.B.: {Fourier Transform Mass Spectrometry Studies of
  Negative Ion Processes}, chap.~11, pp. 175--191.
\newblock American Chemical Society (1987).
\newblock \doi{10.1021/bk-1987-0359.ch011}.
\newblock
  \urlprefix\url{https://pubs.acs.org/doi/abs/10.1021/bk-1987-0359.ch011}

\bibitem{Cavalie2020}
Cavali\'e, T., Venot, O., Yamila, M., Fletcher, L., Wurz, P., Mousis, O.,
  Bounaceur, R., Hue, V., Leconte, J., Dobrijevic, M.: {The deep composition of
  Uranus and Neptune from in situ exploration and thermochemical modeling}.
\newblock Space Science Review  (2020).
\newblock This issue

\bibitem{Colin1979}
Colin, L.: {Encounter with Venus: An Update}.
\newblock Science \textbf{205}(4401), 44--46 (1979).
\newblock \doi{10.1126/science.205.4401.44}

\bibitem{Cotter1992}
Cotter, R.J.: {Time-of-Flight Mass Spectrometry : Instrumentation and
  Applications in Biological Research}.
\newblock American Chemical Society (1992)

\bibitem{Darrach2015}
Darrach, M.R., Madzunkov, S., Schaefer, R., Nikolic, D., Simcic, J., Richard,
  K., Neidholdt, E., Pilinski, M., Jaramillo-Botero, A., Farley, K.: {The Mass
  Analyzer for Real-time Investigation of Neutrals at Europa (MARINE)}.
\newblock 2015 IEEE Aerospace Conference pp. 1--13 (2015)

\bibitem{Dawson1976}
Dawson, P.H.: {Quadrupole Mass Spectrometry and its Applications}.
\newblock Elsevier (1976).
\newblock \doi{https://doi.org/10.1016/C2013-0-04436-2}

\bibitem{Durry2002}
{Durry}, G., {Hauchecorne}, A., {Ovarlez}, J., {Ovarlez}, H., {Pouchet}, I.,
  {Zeninari}, V., {Parvitte}, B.: {In situ Measurement of H$_2$O and CH$_4$
  with Telecommunication Laser Diodes in the Lower Stratosphere: Dehydration
  and Indication of a Tropical Air Intrusion at Mid-Latitudes}.
\newblock Journal of Atmospheric Chemistry \textbf{43}(3), 175--194 (2002).
\newblock \doi{10.1023/A:1020674208207}

\bibitem{Durry2010}
Durry, G., Li, J., Vinogradov, I., Titov, A., Joly, L., Cousin, J.,
  Decarpenterie, T., Amarouche, N., Liu, X., Parvitte, B., Korablev, O.,
  Gerasimov, M., Zéninari, V.: {Near infrared diode laser spectroscopy of
  C$_2$H$_2$, H$_2$O, CO$_2$ and their isotopologues and the application to
  TDLAS, a tunable diode laser spectrometer for the martian PHOBOS-GRUNT space
  mission}.
\newblock Applied Physics B \textbf{99}, 339--351 (2010).
\newblock \doi{10.1007/s00340-010-3924-y}

\bibitem{Foust2019}
Foust, J.: {Europa Clipper passes key review}.
\newblock Space News  (2019).
\newblock
  \urlprefix\url{https://spacenews.com/europa-clipper-passes-key-review/}

\bibitem{Getty2012}
Getty, S.A., Brinckerhoff, W.B., Cornish, T., Ecelberger, S., Floyd, M.:
  Compact two-step laser time-of-flight mass spectrometer for in situ analyses
  of aromatic organics on planetary missions.
\newblock Rapid Communications in Mass Spectrometry \textbf{26}(23), 2786--2790
  (2012).
\newblock \doi{10.1002/rcm.6393}.
\newblock
  \urlprefix\url{https://onlinelibrary.wiley.com/doi/abs/10.1002/rcm.6393}

\bibitem{Glassmeier2007}
{Glassmeier}, K.H., {Boehnhardt}, H., {Koschny}, D., {K{\"u}hrt}, E.,
  {Richter}, I.: {The Rosetta Mission: Flying Towards the Origin of the Solar
  System}.
\newblock Space Science Reviews \textbf{128}(1-4), 1--21 (2007).
\newblock \doi{10.1007/s11214-006-9140-8}

\bibitem{Goesmann2007}
{Goesmann}, F., {Rosenbauer}, H., {Roll}, R., {Szopa}, C., {Raulin}, F.,
  {Sternberg}, R., {Israel}, G., {Meierhenrich}, U., {Thiemann}, W.,
  {Munoz-Caro}, G.: {Cosac, The Cometary Sampling and Composition Experiment on
  Philae}.
\newblock Space Science Review \textbf{128}(1-4), 257--280 (2007).
\newblock \doi{10.1007/s11214-006-9000-6}

\bibitem{Grasset2013}
Grasset, O., Dougherty, M.K., Coustenis, A., Bunce, E.J., Erd, C., Titov, D.,
  Blanc, M., Coates, A., Drossart, P., Fletcher, L.N., Hussmann, H., Jaumann,
  R., Krupp, N., Lebreton, J.P., Prieto-Ballesteros, O., Tortora, P., Tosi, F.,
  Hoolst, T.V.: {JUpiter ICy moons Explorer (JUICE): An ESA mission to orbit
  Ganymede and to characterise the Jupiter system}.
\newblock Planetary and Space Science \textbf{78}, 1--21 (2013).
\newblock \doi{10.1016/j.pss.2012.12.002}

\bibitem{Grinfeld2019}
{Grinfeld}, D., {Stewart}, H., {Skoblin}, M., E., D., {Monastryrsky}, M.,
  {Makarov}, A.: {Space-charge dynamics in Orbitrap mass spectrometers}.
\newblock International Journal of Modern Physics A \textbf{34}(36), 15 (2019).
\newblock \doi{10.1142/S0217751X19420077}.
\newblock 1942007

\bibitem{Haessig2017}
H{\"a}ssig, M., Altwegg, K., Balsiger, H., Berthelier, J., Bieler, A.,
  Calmonte, U., Dhooghe, F., Fiethe, B., Fuselier, S., Gasc, S., Gombosi, T.,
  Le~Roy, L., Luspay-Kuti, A., Mandt, K., Rubin, M., Tzou, C.Y., Wampfler, S.,
  Wurz, P.: {Isotopic composition of CO2 in the coma of
  67P/Churyumov-Gerasimenko measured with ROSINA/DFMS}.
\newblock Astronomy \& Astrophysics \textbf{605}(id.A50) (2017).
\newblock \doi{10.1051/0004-6361/201630140}

\bibitem{Hofer2015}
Hofer, L., Wurz, P., Buch, A., Cabane, M., Coll, P., Coscia, D., Gerasimov, M.,
  Lasi, D., Sapgir, A., Szopa, C., Tulej, M.: {Prototype of the gas
  chromatograph - mass spectrometer to investigate volatile species in the
  lunar soil for the Luna-Resurs mission}.
\newblock Planetary and Space Science \textbf{111}, 126--133 (2015)

\bibitem{Hoffman1972}
{Hoffman}, J.H., {Hodges} R.~R., J., {Evans}, D.E.: {Lunar orbital mass
  spectrometer experiment}.
\newblock Lunar and Planetary Science Conference Proceedings \textbf{3}, 2205
  (1972)

\bibitem{Hoffman1973}
{Hoffman}, J.H., {Hodges} R.~R., J., {Evans}, D.E.: {Lunar Atmospheric
  Composition Results from Apollo 17}.
\newblock In: Lunar and Planetary Science Conference, \emph{Lunar and Planetary
  Science Conference}, vol.~4, p. 376 (1973)

\bibitem{Hofstadter2017}
Hofstadter, M., Simon, A., Reh, K., Elliot, J.: {Ice Giant Mission Study Final
  Report}.
\newblock Tech. rep., {National Aeronautics and Space Administration} and {Jet
  Propulsion Laboratory, California Institute of Technology} (2017).
\newblock JPL D-100520

\bibitem{Hohl1999}
Hohl, M., Wurz, P., Scherer, S., Altwegg, K., Balsiger, H.: {Mass selective
  blanking in a compact multiple reflection time-of-flight mass spectrometer}.
\newblock International Journal of Mass Spectrometry \textbf{188}(3), 189--197
  (1999).
\newblock \doi{https://doi.org/10.1016/S1387-3806(99)00040-8}.
\newblock
  \urlprefix\url{http://www.sciencedirect.com/science/article/pii/S1387380699000408}

\bibitem{Horst2012}
{H{\"o}rst}, S.M., {Yelle}, R.V., {Buch}, A., {Carrasco}, N., {Cernogora}, G.,
  {Dutuit}, O., {Quirico}, E., {Sciamma-O'Brien}, E., {Smith}, M.A., {Somogyi},
  {\'A}., {Szopa}, C., {Thissen}, R., {Vuitton}, V.: {Formation of Amino Acids
  and Nucleotide Bases in a Titan Atmosphere Simulation Experiment}.
\newblock Astrobiology \textbf{12}(9), 809--817 (2012).
\newblock \doi{10.1089/ast.2011.0623}

\bibitem{Hu2005}
Hu, Q., Noll, R.J., Li, H., Makarov, A., Hardman, M., Graham~Cooks, R.: {The
  Orbitrap: a new mass spectrometer}.
\newblock Journal of Mass Spectrometry \textbf{40}(4), 430--443 (2005).
\newblock \doi{10.1002/jms.856}.
\newblock
  \urlprefix\url{https://onlinelibrary.wiley.com/doi/abs/10.1002/jms.856}

\bibitem{Istomin1980}
Istomin, V., Grechnev, K., Kotchnev, V.: {Mass Spectrometer Measurements of the
  Composition of the Lower Atmosphere of Venus}.
\newblock In: M.~Rycroft (ed.) {Space Research}, \emph{{COSPAR Colloquia
  Series}}, vol.~20, pp. 215--218. Pergamon (1980).
\newblock \doi{https://doi.org/10.1016/S0964-2749(13)60044-X}.
\newblock
  \urlprefix\url{http://www.sciencedirect.com/science/article/pii/S096427491360044X}

\bibitem{Johnson1955}
Johnson, C.Y., Meadows, E.B.: First investigation of ambient positive-ion
  composition to 219 km by rocket-borne spectrometer.
\newblock Journal of Geophysical Research (1896-1977) \textbf{60}(2), 193--203
  (1955).
\newblock \doi{10.1029/JZ060i002p00193}.
\newblock
  \urlprefix\url{https://agupubs.onlinelibrary.wiley.com/doi/abs/10.1029/JZ060i002p00193}

\bibitem{Kissel2003}
Kissel, J., Glasmachers, A., Gr{\"u}n, E., Henkel, H., H{\"o}fner, H.,
  Haerendel, G., von Hoerner, H., Hornung, K., Jessberger, E.K., Krueger, F.R.,
  M{\"o}hlmann, D., Greenberg, J.M., Langevin, Y., Sil{\'e}n, J., Brownlee, D.,
  Clark, B.C., Hanner, M.S., Hoerz, F., Sandford, S., Sekanina, Z., Tsou, P.,
  Utterback, N.G., Zolensky, M.E., Heiss, C.: {Cometary and Interstellar Dust
  Analyzer for comet Wild 2}.
\newblock Journal of Geophysical Research: Planets \textbf{108}(E10) (2003).
\newblock \doi{10.1029/2003JE002091}.
\newblock
  \urlprefix\url{https://agupubs.onlinelibrary.wiley.com/doi/abs/10.1029/2003JE002091}

\bibitem{Krankowsky1986}
{Krankowsky}, D., {Lammerzahl}, P., {Herrwerth}, I., {Woweries}, J.,
  {Eberhardt}, P., {Dolder}, U., {Herrmann}, U., {Schulte}, W., {Berthelier},
  J.J., {Illiano}, J.M., {Hodges}, R.R., {Hoffman}, J.H.: {In situ gas and ion
  measurements at comet Halley}.
\newblock Nature \textbf{321}, 326--329 (1986).
\newblock \doi{10.1038/321326a0}

\bibitem{LeBarbu2004}
{Le Barbu}, T., {Vinogradov}, I., {Durry}, G., {Korablev}, O.,
  {Chassefi{\`e}re}, E., {Bertaux}, J.L.: {Tdlas, a diode laser sensor for the
  in situ monitoring of H2O and CO2 isotopes}.
\newblock In: 35th COSPAR Scientific Assembly, vol.~35, p. 2115 (2004)

\bibitem{Lebreton2005}
{Lebreton}, J.P., {Witasse}, O., {Sollazzo}, C., {Blancquaert}, T., {Couzin},
  P., {Schipper}, A.M., {Jones}, J.B., {Matson}, D.L., {Gurvits}, L.I.,
  {Atkinson}, D.H., {Kazeminejad}, B., {P{\'e}rez-Ay{\'u}car}, M.: {An overview
  of the descent and landing of the Huygens probe on Titan}.
\newblock Nature \textbf{438}(7069), 758--764 (2005).
\newblock \doi{10.1038/nature04347}

\bibitem{Li2017}
Li, X., Danell, R.M., Pinnick, V.T., Grubisic, A., [van Amerom], F., Arevalo,
  R.D., Getty, S.A., Brinckerhoff, W.B., Southard, A.E., Gonnsen, Z.D., Adachi,
  T.: Mars organic molecule analyzer (moma) laser desorption/ionization source
  design and performance characterization.
\newblock International Journal of Mass Spectrometry \textbf{422}, 177 -- 187
  (2017).
\newblock \doi{https://doi.org/10.1016/j.ijms.2017.03.010}.
\newblock
  \urlprefix\url{http://www.sciencedirect.com/science/article/pii/S1387380617301513}

\bibitem{Ligterink2019}
Ligterink, N., Riedo, A., Wurz, P., Ehrenfreund, P., Cockell, C., Tulej, M.,
  Grimaudo, V., Lindner, R.: {ORIGIN: a novel and compact Laser Desorption -
  Mass Spectrometry system for sensitive in situ detection of amino acids on
  extraterrestrial surfaces}.
\newblock Nature Science Reports  (2019).
\newblock Submitted

\bibitem{Lorenz2018}
Lorenz, R., Turtle, E., Barnes, J., Trainer, M., Adams, D., Hibbard, K.,
  Sheldon, C., Zacny, K., Peplowski, P., Lawrence, D., Ravine, M., McGee, T.,
  Sotzen, K., MacKenzie, S., Langelaan, J., Schmitz, S., Wolfarth, L., Bedini,
  P.: {Dragonfly: A rotorcraft lander concept for scientific exploration at
  titan}.
\newblock Johns Hopkins APL Technical Digest (Applied Physics Laboratory)
  \textbf{34}(3), 374--387 (2018)

\bibitem{Madzunkov2014}
Madzunkov, S.M., Nikolić, D.: {Accurate Xe Isotope Measurement Using JPL Ion
  Trap}.
\newblock Journal of The American Society for Mass Spectrometry \textbf{25},
  1841--1852 (2014).
\newblock \doi{10.1007/s13361-014-0980-2}

\bibitem{Mahaffy2015}
{Mahaffy}, P.R., {Benna}, M., {King}, T., {Harpold}, D.N., {Arvey}, R.,
  {Barciniak}, M., {Bendt}, M., {Carrigan}, D., {Errigo}, T., {Holmes}, V.,
  {Johnson}, C.S., {Kellogg}, J., {Kimvilakani}, P., {Lefavor}, M.,
  {Hengemihle}, J., {Jaeger}, F., {Lyness}, E., {Maurer}, J., {Melak}, A.,
  {Noreiga}, F., {Noriega}, M., {Patel}, K., {Prats}, B., {Raaen}, E., {Tan},
  F., {Weidner}, E., {Gundersen}, C., {Battel}, S., {Block}, B.P., {Arnett},
  K., {Miller}, R., {Cooper}, C., {Edmonson}, C., {Nolan}, J.T.: {The Neutral
  Gas and Ion Mass Spectrometer on the Mars Atmosphere and Volatile Evolution
  Mission}.
\newblock Space Science Reviews \textbf{195}(1-4), 49--73 (2015).
\newblock \doi{10.1007/s11214-014-0091-1}

\bibitem{Mahaffy2000}
{Mahaffy}, P.R., {Niemann}, H.B., {Alpert}, A., {Atreya}, S.K., {Demick}, J.,
  {Donahue}, T.M., {Harpold}, D.N., {Owen}, T.C.: {Noble gas abundance and
  isotope ratios in the atmosphere of Jupiter from the Galileo Probe Mass
  Spectrometer}.
\newblock Journal of Geophysical Research \textbf{105}(E6), 15061--15072
  (2000).
\newblock \doi{10.1029/1999JE001224}

\bibitem{Mahaffy2012}
Mahaffy, P.R., Webster, C.R., Cabane, M., Conrad, P.G., Coll, P., Atreya, S.K.,
  Arvey, R., Barciniak, M., Benna, M., Bleacher, L., Brinckerhoff, W.B.,
  Eigenbrode, J.L., Carignan, D., Cascia, M., Chalmers, R.A., Dworkin, J.P.,
  Errigo, T., Everson, P., Franz, H., Farley, R., Feng, S., Frazier, G.,
  Freissinet, a., Glavin, D.P., Harpold, D.N., Hawk, D., Holmes, V., Johnson,
  C.S., Jones, A., Jordan, P., Kellogg, J., Lewis, J., Lyness, E., Malespin,
  C.A., Martin, D.K., Maurer, J., McAdam, A.C., McLennan, D., Nolan, T.J.,
  Noriega, M., Pavlov, A.A., Prats, B., Raaen, E., Sheinman, O., Sheppard, D.,
  Smith, J., Stern, J.C., Tan, F., Trainer, M., Ming, D.W., Morris, R.V.,
  Jones, J., Gundersen, C., Steele, A., Wray, J., Botta, O., Leshin, L.A.,
  Owen, T., Battel, S., Jakosky, B.M., Manning, H., Squyres, S.,
  Navarro-Gonz{\'a}lez, R., McKay, C.P., Raulin, F., Sternberg, R., Buch, A.,
  Sorensen, P., Kline-Schoder, R., Coscia, D., Szopa, C., Teinturier, S.,
  Baffes, C., Feldman, J., Flesch, G., Forouhar, S., Garcia, R., Keymeulen, D.,
  Woodward, S., Block, B.P., Arnett, K., Miller, R., Edmonson, C., Gorevan, S.,
  Mumm, E.: {The Sample Analysis at Mars Investigation and Instrument Suite}.
\newblock Space Science Reviews \textbf{170}(1), 401--478 (2012).
\newblock \doi{10.1007/s11214-012-9879-z}

\bibitem{Maiga2018}
Maiga, S.M., Gatica, S.M.: Monolayer adsorption of noble gases on graphene.
\newblock Chemical Physics \textbf{501}, 46--52 (2018).
\newblock \doi{https://doi.org/10.1016/j.chemphys.2017.11.020}.
\newblock
  \urlprefix\url{http://www.sciencedirect.com/science/article/pii/S0301010417306614}

\bibitem{Makarov2000}
Makarov, A.: {Electrostatic axially harmonic orbital trapping: A
  high-performance technique of mass analysis}.
\newblock Analytical Chemistry \textbf{72}(6), 1156--1162 (2000)

\bibitem{Mamyrin1973}
{Mamyrin}, B.A., {Karataev}, V.I., {Shmikk}, D.V., {Zagulin}, V.A.: {The
  mass-reflectron, a new nonmagnetic time-of-flight mass spectrometer with high
  resolution}.
\newblock Soviet Journal of Experimental and Theoretical Physics \textbf{37},
  45 (1973)

\bibitem{March1989}
March, R., Hughes, R.: {Quadrupole Storage Mass Spectrometry}, \emph{Chemical
  Analysis: A Series of Monographs on Analytical Chemistry and Its
  Applications}, vol. 102.
\newblock John Wiley \& Sons (1989).
\newblock \urlprefix\url{https://books.google.ch/books?id=FkYaAQAAMAAJ}

\bibitem{Marshall1990}
Marshall, A., Verdun, V.: {Fourier Transforms in NMR, Optical, and Mass
  Spectrometry}.
\newblock Elsevier (1990)

\bibitem{Mizuno1980}
Mizuno, H.: {Formation of the Giant Planets}.
\newblock Progress of Theoretical Physics \textbf{64}(2), 544--557 (1980).
\newblock \doi{10.1143/PTP.64.544}

\bibitem{Moor1989}
{Moor}, R., {Kopp}, E., {Jenzer}, U., {Ramseyer}, H., {Waelchli}, U., {Arijs},
  E., {Nevejans}, D., {Ingels}, J., {Fussen}, D., {Barassin}, A.: {A double
  focussing mass-spectrometer for simultaneous ion measurements in the
  stratosphere}.
\newblock In: Presented at the 9th ESA Symposium on European Rocket and Balloon
  Programs and Related Research, pp. 3--7 (1989)

\bibitem{Moreno2016}
Moreno-Garc{\'i}a, P., Grimaudo, V., Riedo, A., Tulej, M., Neuland, M.B., Wurz,
  P., Broekmann, P.: {Towards Structural Analysis of Polymeric Contaminants in
  Electrodeposited Cu films}.
\newblock Electrochimica Acta \textbf{199}, 394--402 (2016).
\newblock \doi{https://doi.org/10.1016/j.electacta.2016.03.123}

\bibitem{Mousis2019}
{Mousis}, O., {Atkinson}, D.H., {Ambrosi}, R., {Atreya}, S., {Banfield}, D.,
  {Barabash}, S., {Blanc}, M., {Cavali{\'e}}, T., {Coustenis}, A., {Deleuil},
  M., {Durry}, G., {Ferri}, F., {Fletcher}, L., {Fouchet}, T., {Guillot}, T.,
  {Hartogh}, P., {Hueso}, R., {Hofstadter}, M., {Lebreton}, J.P., {Mandt},
  K.E., {Rauer}, H., {Rannou}, P., {Renard}, J.B., {Sanchez-L{\'a}vega}, A.,
  {Sayanagi}, K., {Simon}, A., {Spilker}, T., {Venkatapathy}, E., {Waite},
  J.H., {Wurz}, P.: {In situ Exploration of the Giant Planets}.
\newblock arXiv e-prints (1908.00917), 1--17 (2019)

\bibitem{Mousis2014}
{Mousis}, O., {Fletcher}, L.N., {Lebreton}, J.P., {Wurz}, P., {Cavali{\'e}},
  T., {Coustenis}, A., {Courtin}, R., {Gautier}, D., {Helled}, R., {Irwin},
  P.G.J., {Morse}, A.D., {Nettelmann}, N., {Marty}, B., {Rousselot}, P.,
  {Venot}, O., {Atkinson}, D.H., {Waite}, J.H., {Reh}, K.R., {Simon}, A.A.,
  {Atreya}, S., {Andr{\'e}}, N., {Blanc}, M., {Daglis}, I.A., {Fischer}, G.,
  {Geppert}, W.D., {Guillot}, T., {Hedman}, M.M., {Hueso}, R., {Lellouch}, E.,
  {Lunine}, J.I., {Murray}, C.D., {O`Donoghue}, J., {Rengel}, M.,
  {S{\'a}nchez-Lavega}, A., {Schmider}, F.X., {Spiga}, A., {Spilker}, T.,
  {Petit}, J.M., {Tiscareno}, M.S., {Ali-Dib}, M., {Altwegg}, K., {Bolton},
  S.J., {Bouquet}, A., {Briois}, C., {Fouchet}, T., {Guerlet}, S., {Kostiuk},
  T., {Lebleu}, D., {Moreno}, R., {Orton}, G.S., {Poncy}, J.: {Scientific
  rationale for Saturn's in situ exploration}.
\newblock Planetary and Space Science \textbf{104}, 29--47 (2014).
\newblock \doi{10.1016/j.pss.2014.09.014}

\bibitem{Niemann1998b}
Niemann, H., Atreya, S., Carignan, G., Donahue, T., Haberman, J., Harpold, D.,
  Hartle, R., Hunten, D., Kasprzak, W., Mahaffy, P., Owen, T., Spencer, N.:
  {Chemical composition measurements of the atmosphere of Jupiter with the
  Galileo Probe mass spectrometer}.
\newblock Advances in Space Research \textbf{21}(11), 1455--1461 (1998).
\newblock \doi{https://doi.org/10.1016/S0273-1177(98)00019-2}.
\newblock
  \urlprefix\url{http://www.sciencedirect.com/science/article/pii/S0273117798000192}

\bibitem{Niemann2002}
{Niemann}, H.B., {Atreya}, S.K., {Bauer}, S.J., {Biemann}, K., {Block}, B.,
  {Carignan}, G.R., {Donahue}, T.M., {Frost}, R.L., {Gautier}, D., {Haberman},
  J.A., {Harpold}, D., {Hunten}, D.M., {Israel}, G., {Lunine}, J.I.,
  {Mauersberger}, K., {Owen}, T.C., {Raulin}, F., {Richards}, J.E., {Way},
  S.H.: {The Gas Chromatograph Mass Spectrometer for the Huygens Probe}.
\newblock Space Science Reviews \textbf{104}(1), 553--591 (2002).
\newblock \doi{10.1023/A:1023680305259}

\bibitem{Niemann2005}
{Niemann}, H.B., {Atreya}, S.K., {Bauer}, S.J., {Carignan}, G.R., {Demick},
  J.E., {Frost}, R.L., {Gautier}, D., {Haberman}, J.A., {Harpold}, D.N.,
  {Hunten}, D.M., {Israel}, G., {Lunine}, J.I., {Kasprzak}, W.T., {Owen}, T.C.,
  {Paulkovich}, M., {Raulin}, F., {Raaen}, E., {Way}, S.H.: {The abundances of
  constituents of Titan's atmosphere from the GCMS instrument on the Huygens
  probe}.
\newblock Nature \textbf{438}(7069), 779--784 (2005).
\newblock \doi{10.1038/nature04122}

\bibitem{Niemann1996}
{Niemann}, H.B., {Atreya}, S.K., {Carignan}, G.R., {Donahue}, T.M., {Haberman},
  J.A., {Harpold}, D.N., {Hartle}, R.E., {Hunten}, D.M., {Kasprzak}, W.T.,
  {Mahaffy}, P.R., {Owen}, T.C., {Spencer}, N.W., {Way}, S.H.: {The Galileo
  Probe Mass Spectrometer: Composition of Jupiter's Atmosphere}.
\newblock Science \textbf{272}(5263), 846--849 (1996).
\newblock \doi{10.1126/science.272.5263.846}

\bibitem{Niemann1998}
{Niemann}, H.B., {Atreya}, S.K., {Carignan}, G.R., {Donahue}, T.M., {Haberman},
  J.A., {Harpold}, D.N., {Hartle}, R.E., {Hunten}, D.M., {Kasprzak}, W.T.,
  {Mahaffy}, P.R., {Owen}, T.C., {Way}, S.H.: {The composition of the Jovian
  atmosphere as determined by the Galileo probe mass spectrometer}.
\newblock Journal of Geophysical Research \textbf{103}(E10), 22831--22846
  (1998).
\newblock \doi{10.1029/98JE01050}

\bibitem{Niemann2010}
Niemann, H.B., Atreya, S.K., Demick, J.E., Gautier, D., Haberman, J.A.,
  Harpold, D.N., Kasprzak, W.T., Lunine, J.I., Owen, T.C., Raulin, F.:
  {Composition of Titan's lower atmosphere and simple surface volatiles as
  measured by the Cassini-Huygens probe gas chromatograph mass spectrometer
  experiment}.
\newblock Journal of Geophysical Research: Planets \textbf{115}(E12) (2010)

\bibitem{Niemann1992}
{Niemann}, H.B., {Harpold}, D.N., {Atreya}, S.K., {Carignan}, G.R., {Hunten},
  D.M., {Owen}, T.C.: {Galileo Probe Mass Spectrometer experiment}.
\newblock Space Science Reviews \textbf{60}(1-4), 111--142 (1992).
\newblock \doi{10.1007/BF00216852}

\bibitem{Nier1947}
Nier, A.O.: {A Mass Spectrometer for Isotope and Gas Analysis}.
\newblock Review of Scientific Instruments \textbf{18}(6), 398--411 (1947).
\newblock \doi{10.1063/1.1740961}

\bibitem{Nier1964}
Nier, A.O., Hoffman, J.H., Johnson, C.Y., Holmes, J.C.: {Neutral constituents
  of the upper atmosphere: The minor peaks observed in a mass spectrometer}.
\newblock Journal of Geophysical Research (1896-1977) \textbf{69}(21),
  4629--4636 (1964).
\newblock \doi{10.1029/JZ069i021p04629}.
\newblock
  \urlprefix\url{https://agupubs.onlinelibrary.wiley.com/doi/abs/10.1029/JZ069i021p04629}

\bibitem{Nier1977}
Nier, A.O., McElroy, M.B.: {Composition and structure of Mars' Upper
  atmosphere: Results from the neutral mass spectrometers on Viking 1 and 2}.
\newblock Journal of Geophysical Research \textbf{82}(28), 4341--4349 (1977).
\newblock \doi{10.1029/JS082i028p04341}.
\newblock
  \urlprefix\url{https://agupubs.onlinelibrary.wiley.com/doi/abs/10.1029/JS082i028p04341}

\bibitem{Nikolaev2011}
Nikolaev, E.N., Boldin, I.A., Jertz, R., Baykut, G.: {Initial Experimental
  Characterization of a New Ultra-High Resolution FTICR Cell with Dynamic
  Harmonization}.
\newblock Journal of The American Society for Mass Spectrometry \textbf{22},
  1125--1133 (2011).
\newblock \doi{10.1007/s13361-011-0125-9}

\bibitem{Nikolic2019}
{Nikoli{\'c}}, {Madzunkov}, {Darrach}: {Response of QIT-MS to Noble Gas
  Isotopic Ratios in a Simulated Venus Flyby}.
\newblock Atmosphere \textbf{10}(5), 232 (2019).
\newblock \doi{10.3390/atmos10050232}

\bibitem{Okumura2004}
Okumura, D., Toyoda, M., Ishihara, M., Katakuse, I.: A compact sector-type
  multi-turn time-of-flight mass spectrometer `multum ii'.
\newblock Nuclear Instruments and Methods in Physics Research Section A:
  Accelerators, Spectrometers, Detectors and Associated Equipment
  \textbf{519}(1), 331--337 (2004).
\newblock \doi{https://doi.org/10.1016/j.nima.2003.11.249}.
\newblock
  \urlprefix\url{http://www.sciencedirect.com/science/article/pii/S0168900203030389}.
\newblock Proceedings of the Sixth International Conference on Charged Particle
  Optics

\bibitem{Orsini2010}
{Orsini}, S., {Livi}, S., {Torkar}, K., {Barabash}, S., {Milillo}, A., {Wurz},
  P., {di Lellis}, A.M., {Kallio}, E., {SERENA Team}: {SERENA: A suite of four
  instruments (ELENA, STROFIO, PICAM and MIPA) on board BepiColombo-MPO for
  particle detection in the Hermean environment}.
\newblock Planetary and Space Science \textbf{58}(1-2), 166--181 (2010).
\newblock \doi{10.1016/j.pss.2008.09.012}

\bibitem{Palmer2001}
Palmer, P.T., Limero, T.F.: {Mass spectrometry in the U.S. space program: Past,
  present, and future}.
\newblock Journal of the American Society for Mass Spectrometry \textbf{12}(6),
  656--675 (2001).
\newblock \doi{10.1021/jasms.8b01630}

\bibitem{Pappalardo2013}
Pappalardo, R., Vance, S., Bagenal, F., Bills, B., Blaney, D., Blankenship, D.,
  Brinckerhoff, W., Connerney, J., Hand, K., Hoehler, T., Leisner, J., Kurth,
  W., McGrath, M., Mellon, M., Moore, J., Patterson, G., Prockter, L., Senske,
  D., Schmidt, B., Shock, E., Smith, D., Soderlund, K.: {Science Potential from
  a Europa Lander}.
\newblock Astrobiology \textbf{13}(8), 740--773 (2013).
\newblock \doi{10.1089/ast.2013.1003}.
\newblock \urlprefix\url{https://doi.org/10.1089/ast.2013.1003}.
\newblock PMID: 23924246

\bibitem{Pollack1996}
Pollack, J.B., Hubickyj, O., Bodenheimer, P., Lissauer, J.J., Podolak, M.,
  Greenzweig, Y.: {Formation of the Giant Planets by Concurrent Accretion of
  Solids and Gas}.
\newblock Icarus \textbf{124}(1), 62--85 (1996).
\newblock \doi{10.1006/icar.1996.0190}

\bibitem{Poole2012}
Poole, C.: {Gas Chromatography}.
\newblock Handbooks in Separation Science. Elsevier Science (2012).
\newblock \urlprefix\url{https://books.google.ch/books?id=O77061hwfd4C}

\bibitem{Riedo2016}
Riedo, A., Grimaudo, V., Moreno-Garc{\'i}a, P., Neuland, M.B., Tulej, M.,
  Broekmann, P., Wurz, P.: {Laser Ablation/Ionisation Mass Spectrometry:
  Sensitive and Quantitative Chemical Depth Profiling of Solid Materials}.
\newblock CHIMIA International Journal for Chemistry \textbf{70}(4), 268--273
  (2016).
\newblock \doi{doi:10.2533/chimia.2016.268}

\bibitem{Riedo2015}
Riedo, A., Grimaudo, V., Moreno-Garc{\'i}a, P., Neuland, M.B., Tulej, M., Wurz,
  P., Broekmann, P.: {High depth-resolution laser ablation chemical analysis of
  additive-assisted Cu electroplating for microchip architectures}.
\newblock Journal of Analytical Atomic Spectrometry \textbf{30}, 2371--2374
  (2015).
\newblock \doi{10.1039/C5JA00295H}

\bibitem{Riedo2019}
Riedo, A., de~Koning, C., Stevens, A., McDonald, A., Lopez, A.C., Tulej, M.,
  Wurz, P., Cockell, C., Ehrenfreund, P.: {The detection of microbes in Martian
  mudstone analogues using laser ablation ionization mass spectrometry at high
  spatial resolution}.
\newblock Astrobiology  (2019).
\newblock Submitted

\bibitem{Rubin2018}
Rubin, M., Altwegg, K., Balsiger, H., Bar-Nun, A., Berthelier, J.J., Briois,
  C., Calmonte, U., Combi, M., De~Keyser, J., Fiethe, B., Fuselier, S., Gasc,
  S., Gombosi, T., Hansen, K., Kopp, E., Korth, A., Laufer, D., Le~Roy, L.,
  Mall, U., Marty, B., Mousis, O., Owen, T., R{\`e}me, H., S{\'e}mon, T., Tzou,
  C.Y., Waite, J., Wurz, P.: {Krypton isotopes and noble gas abundances in the
  coma of comet 67P/Churyumov-Gerasimenko}.
\newblock Science Advances \textbf{4}(eaar6297) (2018).
\newblock \doi{10.1126/sciadv.aar6297}

\bibitem{Rubin2019}
{Rubin}, M., {Bekaert}, D.V., {Broadley}, M.W., {Drozdovskaya}, M.N.,
  {Wampfler}, S.F.: {Volatile Species in Comet 67P/Churyumov-Gerasimenko:
  Investigating the Link from the ISM to the Terrestrial Planets}.
\newblock ACS Earth and Space Chemistry \textbf{3}(9), 1792--1811 (2019).
\newblock \doi{10.1021/acsearthspacechem.9b00096}

\bibitem{Rushneck1978}
Rushneck, D.R., Diaz, A.V., Howarth, D.W., Rampacek, J., Olson, K.W., Dencker,
  W.D., Smith, P., McDavid, L., Tomassian, A., Harris, M., Bulota, K., Biemann,
  K., LaFleur, A.L., Biller, J.E., Owen, T.: {Viking gas chromatograph-mass
  spectrometer}.
\newblock Review of Scientific Instruments \textbf{49}(6), 817--834 (1978).
\newblock \doi{10.1063/1.1135623}

\bibitem{Scherer2006}
Scherer, S., Altwegg, K., Balsiger, H., Fischer, J., J{\"a}ckel, A., Korth, A.,
  Mildner, M., Piazza, D., Reme, H., Wurz, P.: {A novel principle for an ion
  mirror design in time-of-flight mass spectrometry}.
\newblock International Journal of Mass Spectrometry \textbf{251}(1), 73--81
  (2006).
\newblock \doi{https://doi.org/10.1016/j.ijms.2006.01.025}.
\newblock
  \urlprefix\url{http://www.sciencedirect.com/science/article/pii/S1387380606000510}

\bibitem{Schletti2001}
{Schletti}, R., {Wurz}, P., {Scherer}, S., {Siegmund}, O.H.: {Fast microchannel
  plate detector with an impedance matched anode in suspended substrate
  technology}.
\newblock Review of Scientific Instruments \textbf{72}(3), 1634--1639 (2001).
\newblock \doi{10.1063/1.1344601}

\bibitem{Schroeder2019a}
Schroeder~I, I., Altwegg, K., Balsiger., H., Berthelier, J.J., Combi, M.,
  De~Keyser, J., Fiethe, B., Fuselier, S., Gombosi, T., Hansen, K., Rubin, M.,
  Shou, Y., Tenishev, V., S{\'e}mon, T., Wampfler, S., Wurz, P.: {A comparison
  between the two lobes of comet 67P / Churyumov-Gerasimenko based on D/H
  ratios in H$_2$O measured with the Rosetta / ROSINA DFMS}.
\newblock Monthly Notices of the Royal Astronomical Society \textbf{489},
  4734--4740 (2019).
\newblock \doi{10.1093/mnras/stz2482}

\bibitem{Schroeder2019b}
Schroeder~I, I., Altwegg, K., Balsiger., H., Berthelier, J.J., De~Keyser, J.,
  Fiethe, B., Fuselier, S., Gasc, S., Gombosi, T., Rubin, M., S{\'e}mon, T.,
  Tzou C.-Y.and~Wampfler, S., Wurz, P.: {The $^{16}{\rm O}/^{18}{\rm O}$ Ratio
  in Water in the Coma of Comet 67P/Churyumov-Gerasimenko measured with the
  Rosetta/ROSINA Double Focusing Mass Spectrometer}.
\newblock Astronomy \& Astrophysics \textbf{630}(A29) (2019).
\newblock \doi{10.1051/0004-6361/201833806}

\bibitem{Schulz2006}
Schulz, R., Benkhoff, J.: {BepiColombo: Payload and mission updates}.
\newblock Advances in Space Research \textbf{38}(4), 572--577 (2006).
\newblock \doi{https://doi.org/10.1016/j.asr.2005.05.084}.
\newblock
  \urlprefix\url{http://www.sciencedirect.com/science/article/pii/S0273117705007003}.
\newblock Mercury, Mars and Saturn

\bibitem{Selliez2019}
Selliez, L., Briois, C., Carrasco, N., Thirkell, L., Thissen, R., Ito, M.,
  Orthous-Daunay, F.R., Chalumeau, G., Colin, F., Cottin, H., Engrand, C.,
  Flandinet, L., Fray, N., Gaubicher, B., Grand, N., Lebreton, J.P., Makarov,
  A., Ruocco, S., Szopa, C., Vuitton, V., Zapf, P.: Identification of organic
  molecules with a laboratory prototype based on the laser
  ablation-cosmorbitrap.
\newblock Planetary and Space Science \textbf{170}, 42 -- 51 (2019).
\newblock \doi{https://doi.org/10.1016/j.pss.2019.03.003}.
\newblock
  \urlprefix\url{http://www.sciencedirect.com/science/article/pii/S0032063318302666}

\bibitem{Shimma2012}
Shimma, S., Toyoda, M.: {Miniaturized Mass Spectrometer in Analysis of
  Greenhouse Gases: The Performance and Possibilities}.
\newblock In: G.~Liu (ed.) {Greenhouse Gases}, chap.~11. IntechOpen, Rijeka
  (2012).
\newblock \doi{10.5772/33815}

\bibitem{Simon2020}
Simon, A.A., Fletcher, L.N., Arridge, C., Atkinson, D., Coustensis, A., Ferri,
  F., Hofstadter, M., Masters, A., Mousis, O., Reh, K., Turrini, D., Witasse,
  O.: {Recent mission concepts for in situ exploration of the Ice Giants}.
\newblock {Space Science Review}  (2020).
\newblock This issue

\bibitem{Smith2018}
Smith, D.F., Podgorski, D.C., Rodgers, R.P., Blakney, G.T., Hendrickson, C.L.:
  21 tesla ft-icr mass spectrometer for ultrahigh-resolution analysis of
  complex organic mixtures.
\newblock Analytical Chemistry \textbf{90}(3), 2041--2047 (2018).
\newblock \doi{10.1021/acs.analchem.7b04159}.
\newblock \urlprefix\url{https://doi.org/10.1021/acs.analchem.7b04159}.
\newblock PMID: 29303558

\bibitem{Snodgrass2019}
Snodgrass, C., Jones, G.H.: {TI - The European Space Agency's Comet Interceptor
  lies in wait}.
\newblock {Nature Communications} \textbf{10} (2019).
\newblock \doi{10.1038/s41467-019-13470-1}

\bibitem{Toyoda2003}
{Toyoda}, M., {Okumura}, D., {Ishihara}, M., {Katakuse}, I.: {Multi-turn
  time-of-flight mass spectrometers with electrostatic sectors}.
\newblock {Journal of Mass Spectrometry} \textbf{38}(11), 1125--1142 (2003).
\newblock \doi{10.1002/jms.546}

\bibitem{Vinogradov1971}
Vinogradov, A.P., Surkov, Y.A., Andreichikov, B.M., Kalinkina, O.M.,
  Grechischeva, I.M.: {The chemical composition of the atmosphere of Venus}.
\newblock Symposium - International Astronomical Union \textbf{40}, 3--16
  (1971).
\newblock \doi{10.1017/S0074180900102529}

\bibitem{vonZahn1992}
Von~Zahn, U., Hunten, D.: {The Jupiter helium interferometer experiment on the
  Galileo Entry Probe}.
\newblock Space Science Reviews \textbf{60}(1--4), 263--281 (1992).
\newblock \doi{10.1007/BF00216857}

\bibitem{vonZahn1998}
Von~Zahn, U., Hunten, D., Lehmacher, G.: {Helium in Jupiter's atmosphere:
  Results from the Galileo probe Helium Interferometer Experiment}.
\newblock {Journal of Geophysical Research E: Planets} \textbf{103}(E10),
  22815--22829 (1998).
\newblock \doi{10.1029/98JE00695}

\bibitem{Waller2019}
Waller, S.E., Belousov, A., Kidd, R.D., Nikoli{\'c}, D., Madzunkov, S.M.,
  Wiley, J.S., Darrach, M.R.: {Chemical Ionization Mass Spectrometry:
  Applications for the In Situ Measurement of Nonvolatile Organics at Ocean
  Worlds}.
\newblock {Astrobiology} \textbf{19}(10) (2019).
\newblock \doi{10.1089/ast.2018.1961}

\bibitem{Webster2013}
Webster, C.R., Mahaffy, P.R., Flesch, G.J., Niles, P.B., Jones, J.H., Leshin,
  L.A., Atreya, S.K., Stern, J.C., Christensen, L.E., Owen, T., Franz, H.,
  Pepin, R.O., Steele, A.: {Isotope Ratios of H, C, and O in CO2 and H2O of the
  Martian Atmosphere}.
\newblock Science \textbf{341}(6143), 260--263 (2013).
\newblock \doi{10.1126/science.1237961}.
\newblock \urlprefix\url{https://science.sciencemag.org/content/341/6143/260}

\bibitem{Wong2004}
Wong, M.H., Mahaffy, P.R., Atreya, S.K., Niemann, H.B., Owen, T.C.: {Updated
  Galileo probe mass spectrometer measurements of carbon, oxygen, nitrogen, and
  sulfur on Jupiter}.
\newblock Icarus \textbf{171}(1), 153--170 (2004).
\newblock \doi{https://doi.org/10.1016/j.icarus.2004.04.010}.
\newblock
  \urlprefix\url{http://www.sciencedirect.com/science/article/pii/S0019103504001393}

\bibitem{Wright2007}
{Wright}, I.P., {Barber}, S.J., {Morgan}, G.H., {Morse}, A.D., {Sheridan}, S.,
  {Andrews}, D.J., {Maynard}, J., {Yau}, D., {Evans}, S.T., {Leese}, M.R.,
  {Zarnecki}, J.C., {Kent}, B.J., {Waltham}, N.R., {Whalley}, M.S., {Heys}, S.,
  {Drummond}, D.L., {Edeson}, R.L., {Sawyer}, E.C., {Turner}, R.F.,
  {Pillinger}, C.T.: {Ptolemy---an Instrument to Measure Stable Isotopic Ratios
  of Key Volatiles on a Cometary Nucleus}.
\newblock Space Science Review \textbf{128}(1-4), 363--381 (2007).
\newblock \doi{10.1007/s11214-006-9001-5}

\bibitem{Wurz2012}
{Wurz}, P., {Abplanalp}, D., {Tulej}, M., {Lammer}, H.: {A neutral gas mass
  spectrometer for the investigation of lunar volatiles}.
\newblock Planetary Space Science \textbf{74}(1), 264--269 (2012).
\newblock \doi{10.1016/j.pss.2012.05.016}

\bibitem{Wurz2015}
{Wurz}, P., {Rubin}, M., {Altwegg}, K., {Balsiger}, H., {Berthelier}, J.J.,
  {Bieler}, A., {Calmonte}, U., {De Keyser}, J., {Fiethe}, B., {Fuselier},
  S.A., {Galli}, A., {Gasc}, S., {Gombosi}, T.I., {J{\"a}ckel}, A., {Le Roy},
  L., {Mall}, U.A., {R{\`e}me}, H., {Tenishev}, V., {Tzou}, C.Y.: {Solar wind
  sputtering of dust on the surface of 67P/Churyumov-Gerasimenko}.
\newblock Astronomy and Astrophysics \textbf{583}, A22 (2015).
\newblock \doi{10.1051/0004-6361/201525980}

\end{thebibliography}


\end{document}